\documentclass[a4paper,10pt,twoside]{cpc-hepnp}
\usepackage{CJK,upgreek,fancyhdr}
\usepackage{multicol}
\usepackage{graphicx}
\usepackage{booktabs}
\usepackage{amssymb,bm,mathrsfs,bbm,amscd}
\usepackage[tbtags]{amsmath}
\usepackage{lastpage}
\usepackage{subcaption}

\begin{document}
\begin{CJK*}{GB}{gbsn}

\fancyhead[c]{\small Chinese Physics C~~~Vol. xx, No. x (201x) xxxxxx}
\fancyfoot[C]{\small 010201-\thepage}

\footnotetext[0]{Received 31 June 2015}

\title{Spectroscopic Investigation of Light Strange S=-1 $\Lambda$, $\Sigma$ and S=-2 $\Xi$ Baryons}

\author{%
      Chandni Menapara, \email{chandni.menapara@gmail.com}%
and
 Ajay Kumar Rai
}
\maketitle

\address{%
Department of Physics, Sardar Vallabhbhai National Institute of Technology, Surat-395007, Gujarat, India\\
% $^2$ {\bf Example}: Institute of High Energy Physics, Chinese Academy of Sciences, Beijing 100049, China\\
}

\begin{abstract}
{\bf The present study is dedicated to light-strange $\Lambda$ with strangeness S=-1, isospin I=0; $\Sigma$ with S=-1, I=1 and $\Xi$ baryon with  S=-2 and $I=\frac{1}{2}$. In this article, hypercentral Constituent Quark Model with linear confining potential has been employed along with first order correction term to obtain the resonance masses for nearly upto 4 GeV. The calculated states include 1S-5S, 1P-4P, 1D-3D, 1F-2F  and 1G (in few case) along with all the possible spin-parity assignments. Regge Trajectories have been explored for the linearly of the calculated masses for $(n,M^{2})$ and $(J,M^{2})$ respectively. Magnetic moments have been intensively studied for ground state spin $\frac{1}{2}$ and $\frac{3}{2}$, in addition to the configuration mixing for first negative parity state for $\Xi$. Lastly, transition magnetic moment and radiative decay width have been presented. } 
\end{abstract}


\begin{pacs}
1 -- 3 PACS codes (Physics and Astronomy Classification Scheme, https://publishing.aip.org/publishing/pacs/pacs-2010-regular-edition/)
\end{pacs}

\footnotetext[0]{\hspace*{-3mm}\raisebox{0.3ex}{$\scriptstyle\copyright$}2013
Chinese Physical Society and the Institute of High Energy Physics of the Chinese Academy of Sciences and the Institute of Modern Physics of the Chinese Academy of Sciences and IOP Publishing Ltd}%

\begin{multicols}{2}

\section{Introduction}
\label{intro}
The objective for the study of hadrons is to reveal the possible degrees of freedom responsible for the way a given system appears. The quark confinement and asymptotic freedom has been the starting point of any theoretical and phenomenological study to understand quark dynamics. Hadron spectroscopy attempts to explore the excited mass spectrum along with the multiplet structure as well as  spin-parity assignments. The light quark baryons form the basis of octet and decuplet ranging from strangeness S=0 to S=-3 based on the symmetric, asymmetric and mixed-symmetric flavor-spin combinations. 
$$ 3 \otimes 3 \otimes 3 = 1^{A} \oplus 8^{M} \oplus 8^{M} \oplus 10^{S} $$
The presence of strange quark in a baryon draws attention because of the fact that it is a bit heavier compared to u and d quarks whereas considerably light compared to c and b quarks. Particularly the strangeness S=-2 $\Xi$ baryons have not been observed experimentally like other light sector baryons \cite{pdg} as depicted in table \ref{tab:pdg}. The limited observations in the $\Xi$ baryon group owes to the fact that they are produced only as the final state in a process in addition to considerably small cross-sections \cite{pervin}. {\bf Unlike the $\Xi$ baryons, $\Sigma$ and $\Lambda$ with S=-1 have quite a number of experimentally established states.}
\\\\
The present article is dedicated to the study of $\Lambda$, $\Sigma$ and $\Xi$ baryons. Cascade baryons appear with isospin $I=\frac{1}{2}$ in the octet ($J=\frac{1}{2}$) and decuplet ($J=\frac{3}{2}$) as $\Xi$ and $\Xi^{*}$ respectively. The quark combination is uss for $\Xi^{0}$ and dss for $\Xi^{-}$.  \\
For mixed symmetry flavour wave-function in octet group,
\begin{equation*}
\Xi^{0} \rightarrow \quad \frac{1}{\sqrt{6}}(sus+uss-2ssu)
\end{equation*}
\begin{equation*}
\Xi^{-} \rightarrow \quad \frac{1}{\sqrt{6}}(dss+sds-2ssd)
\end{equation*}
\begin{equation*}
\Xi^{0} \rightarrow \quad \frac{1}{\sqrt{2}}(uss-sus)
\end{equation*}
\begin{equation*}
\Xi^{-} \rightarrow \quad \frac{1}{\sqrt{2}}(dss-sds)
\end{equation*}
For symmetric flavour wave-function in decuplet group, 
\begin{equation*}
\Xi^{*0} \rightarrow \quad \frac{1}{\sqrt{3}}(uss+sus+ssu)
\end{equation*}
\begin{equation*}
\Xi^{*-} \rightarrow \quad \frac{1}{\sqrt{3}}(dss+sds+ssd)
\end{equation*}

{\bf Similarly the $\Sigma$ baryon with u, d and s constituent quarks have place in octet and decuplet with three possible combinations as uus, uds and dds respectively. The $\Lambda$ baryon appearing in octet as I=0 has uds quark content however, it differs from $\Sigma^{0}$ based on the wave-function. } \\
\begin{table*}
\centering
\caption{The $\Xi$ listed by Particle Data Group (PDG\cite{pdg})}
\label{tab:pdg}
\begin{tabular}{ccc}
\hline
State & $J^{P}$ & Status \\
\hline
$\Xi^{0}$(1314) & $\frac{1}{2}^{+}$ & **** \\
$\Xi^{-}$(1321) & $\frac{1}{2}^{+}$ & **** \\
$\Xi$(1530) & $\frac{3}{2}^{+}$ & **** \\
$\Xi$(1620) & & * \\
$\Xi$(1690) & & *** \\
$\Xi$(1820) & $\frac{3}{2}^{-}$ & *** \\
$\Xi$(1950) & & *** \\
$\Xi$(2030) & $\frac{5}{2}^{?}$ & *** \\
$\Xi$(2120) & & * \\
$\Xi$(2250) & & ** \\
$\Xi$(2370) & & ** \\
$\Xi$(2500) & & * \\
\hline
\end{tabular}
\end{table*}

Experimental facilities across the world have been striving towards the study of strange hyperons. A recent study at CERN by ALICE Collaboration has established an attractive interaction of proton and $\Xi^{-}$ \cite{alice}.
Activities in measuring weak decays of $\Xi$ hyperon were reported by the KTeV Collaboration \cite{ktev} and by the NA48/1 Collaboration \cite{na48} as well as BABAR Collaboration has been carrying out extensive studies \cite{babar}. The photoproduction of $\Xi$ has been observed by CLAS detector at Jefferson Lab \cite{clas}. ALso, recently JLab has proposed to explore the strange hyperon spectroscopy through secondary KL beam alongwith GlueX experiment \cite{jlab} and the findings shall expectantly give new directions and understanding of strange hyperons $\Sigma$, $\Lambda$ and $\Xi$. The BESIII Collaboration observed $\Xi$(1530) in baryon-antibaryon pair from charmonium decay \cite{bes}. The upcoming experimental facility PANDA at FAIR-GSI has been highly expected to establish the whole spectrum of hyperons through proton-antiproton collisions \cite{panda}. A $\Xi$ dedicated study has been undertaken for mass, widths and decay modes by one of the PANDA group \cite{jputz,barruca}. 
\\\\
{\bf In case of $\Sigma$ and $\Lambda$ baryons, all the properties are not known completely. Most of the data for strange baryons have been based from earlier studies from bubble chamber for $K^{-}$ reactions. The $\Lambda$(1405) with $J^{P}=\frac{1}{2}^{-}$ is still a mysterious state in the lambda spectrum. This state is lower than the non-strange counterpart $N^{*}$(1535). Recent studies have attempted to understand this state as some hadronic molecular \cite{mai, jido}. JPAC and Osaka-ANL group have applied coupled channel approach to study this dynamics \cite{jpac, osaka}. Also, two pole structure of $\Lambda$(1405) is analyzed using chiral effective field theory \cite{ren}. E. Klempt {\it et al.} has extensively reviewed the $\Lambda$ and $\Sigma$ hyperon spectrum based on experimental and theoretical studies focusing on all the known states of the spectrum \cite{klempt20}. These spectrums are studied through photoproduction off the proton in ref \cite{kim}. The study of strangeness S=-1, -2 becomes more interesting not only in high energy arena but also in astrophysical bodies as in neutron stars \cite{jung}. }\\\\

The excited states of hyperons have been investigated using phenomenological as well as theoretical approaches. Various models have attempted to reproduce the octet and decuplet ground states and a range of excited states. A recent review has summarized in details few of the states of strange baryons spectrum with theoretical and experimental aspects \cite{hyodo}. L. Xiao et al. has intensively studied the strong decays of $\Xi$ under Chiral quark model which may assist in determining possible spin-parity of a given state in strong decay \cite{xiao}. Some of the models have been summarized briefly in Section 3 which ranges from relativistic appraoch \cite{faustov}, instanton induced model \cite{loring}, CI model \cite{capstick}, algebraic model \cite{bijker}, Skyrme model \cite{oh}, etc. The present work has been lead on the phenomenological non-relativistic hypercentral Constituent Quark Model \cite{giannini}. Similar studies have been carried out for nucleons and delta baryons earlier which serve as a key to proceed towards exploring the strange baryons \cite{cpc,z19}.\\\\
This paper is organized as follows: after the briefing of hadron spectroscopy and various experimental and theoretical approaches, section 2 deals with the background of the model used. Section 3 describes the results for resonance masses alongwith the detailed discussion about the data presented in respective tables. The later section 4 represents the Regge trajectories and deductions based on them. Section 5 is dedicated to magnetic moments of isospin states of $\Xi$, $\Sigma$ and $\Lambda$ based on their effective masses. Section 6 focuses on the transition magnetic moment and radiative decay widths and lastly the conclusion.

\section{Theoretical Background}
\label{sec:1}
The Constituent Quark Model (CQM) is based on a simple assumption of baryon as a system of three quarks (or anti-quarks)interacting by some potential which ultimately may provide some quantitative description of baryonic properties. It is obvious that the QCD quark masses are considerably smaller than constituent quark masses however this large mass parametrizes all the other effects within a baryon.  Thus, the CQMs have been employed in various studies through various modifications in non-relativistic or semi-relativistic approaches. \\\\
The hypercentral Constituent Quark Model (hCQM) has been employed for the present study which is a non relativistic approach \cite{rai}. It undertakes the baryon as a confined system of three quarks wherein the potential is hypercentral one. The dynamics of three body system are taken care of using the Jacobi coordinates introduced as $\rho$ and $\lambda$ reducing it to two body parameters \cite{ferraris}. 
\begin{subequations}
\begin{align}
 {\bf \rho} = \frac{1}{\sqrt{2}}({\bf r_{1}} -{\bf r_{2}}) \\
{\bf \lambda} = \frac{(m_{1}{\bf r_{1}} + m_{2}{\bf r_{2}} - (m_{1}+m_{2}){\bf r_{3}})}{\sqrt{m_{1}^{2} + m_{2}^{2} + (m_{1}+m_{2})^{2}}} 
\end{align}
\end{subequations}
The hyperradius and hyperangle are defined as 
\begin{equation}
x = \sqrt{{\bf \rho^{2}} + {\bf \lambda^{2}}}
; \; \; \xi = arctan(\frac{\rho}{\lambda})
\end{equation} 
The the Hamiltonian of the system is written with potential term solely depending on hyperradius x of three body systems
\begin{equation}
H = \frac{P^{2}}{2m} + V^{0}(x)  + V_{SD}(x)
\end{equation}
where $m=\frac{2m_{\rho}m_{\lambda}}{m_{\rho}+m_{\lambda}}$ is the reduced mass.
Thus the  hyper-radial part of the wave-function as determined by hypercentral Schrodinger equation is \cite{giannini}
\begin{equation}
\left[\frac{d^{2}}{dx^{2}} + \frac{5}{x}\frac{d}{dx} - \frac{\gamma(\gamma +4)}{x^{2}}\right]\psi(x) = -2m[E-V(x)]\psi(x)
\end{equation}
Here $\gamma$ replaces the angular momentum quantum number by the relation as $l(l+1) \rightarrow \frac{15}{4} + \gamma (\gamma + 4)$.
The choice of hypercentral potential narrows down to hyperCoulomb one i.e. $-\frac{\tau}{x} $. The confinement term here is chosen to be of linear nature.
\begin{equation}
 V^{0}(x) = -\frac{\tau}{x} + \alpha x
\end{equation}
Here, $\tau = \frac{2}{3}\alpha_{s}$ with $\alpha_{s}$ being running coupling constant and $\alpha$ is a parameter based on the fitting of ground state for a given system.

The $V_{SD}(x)$ accounts for the spin-dependent terms leading to hyperfine interactions. 
\begin{equation}
\begin{split}
V_{SD}(x) = V_{SS}(x)({\bf S_{\rho}\cdot S_{\lambda}}) +  V_{\gamma S}(x)({\bf \gamma \cdot S}) \\
+ V_{T}\times [S^{2}- \frac{3({\bf S\cdot x})({\bf S\cdot x})}{x^{2}}]
\end{split}
\end{equation}
where $V_{SS}(x)$, $V_{\gamma S}(x)$ and $V_{T}(x)$ are spin-spin, spin-orbit and tensor terms respectively \cite{voloshin}.
In the present study, first order correction term to potential with $\frac{1}{m}$ dependence has also been added as $\frac{1}{m}V^{1}(x)$ \cite{zalak}. 
\begin{equation}
V^{1}(x)= -C_{F}C_{A}\frac{\alpha_{s}^{2}}{4x^{2}}
\end{equation}
where $C_{F}$ and $C_{A}$ are Casimir elements of fundamental and adjoint representation. \\
The resonance masses have been obtained with and without the firt order correction term. The constituent quark masses have been taken to be $m_{u}=m_{d}=0.290$ MeV and $m_{s}=0.500$ MeV. Mathematica Notebook has been employed for numerical solutions \cite{lucha}.

\section{Results and Discussion for the Resonance Mass Spectra}
In the present work, the resonance masses are calculated for radial and orbital states from 1S-5S, 1P-4P, 1D-3D and 1F-2F. Also, all the possible spin-parity configuration for each state with $S=\frac{1}{2}$ and $S=\frac{3}{2}$ have been considered and calculated the respective contribution based on the model discussed in section 2 including with and without first order corrections. For S-state, possible total angular momentum and parity are $\frac{1}{2}^{+}$ and $\frac{3}{2}^{+}$, for P-state the range goes from $\frac{1}{2}^{-}$ to $\frac{5}{2}^{-}$, for D-state the range is $\frac{3}{2}^{+}$ to $\frac{7}{2}^{+}$ and for F-state, it is $\frac{3}{2}^{-}$ to $\frac{9}{2}^{-}$. Only few experimentally established states with four star status available are mentioned in respective tables. In the tables, $Mass_{cal}1$ and $Mass_{cal}2$ represents resonance masses without and with first order correction respectively in the units of MeV. \\

\begin{table*}
\centering
\caption{S-wave of $\Xi$ baryon (in MeV)}
\label{tab:s-wave}
\begin{tabular}{ccccc}
\hline
State & $J^{P}$ &  $Mass_{cal}$1 & $Mass_{cal}$2 & $Mass_{exp}$ \cite{pdg} \\
\hline
1S & $\frac{1}{2}^{+}$ & 1322 & 1321 & 1321 \\
 & $\frac{3}{2}^{+}$ & 1531 & 1524 & 1532 \\
2S & $\frac{1}{2}^{+}$ & 1884 & 1891 &  \\
  & $\frac{3}{2}^{+}$ & 1971 & 1964 & \\
3S & $\frac{1}{2}^{+}$ & 2361 & 2372 & \\
 & $\frac{3}{2}^{+}$ & 2457 & 2459 & \\
4S & $\frac{1}{2}^{+}$ & 2935 & 2954 & \\
 & $\frac{3}{2}^{+}$ & 3029 & 3041 &  \\ 
5S & $\frac{1}{2}^{+}$ & 3591 & 3620 & \\
 & $\frac{3}{2}^{+}$ & 3679 & 3702 & \\
\hline
\end{tabular}
\end{table*}

\begin{table*}
\centering
\caption{P-wave of $\Xi$ baryon (in MeV)}
\label{tab:p-wave}
\begin{tabular}{ccccc}
\hline
State & $J^{P}$ & $Mass_{cal}$1 & $Mass_{cal}$2 & $Mass_{exp}$\cite{pdg} \\
\hline
$1^{2}P_{1/2}$ & $\frac{1}{2}^{-}$ & 1886 & 1889 & \\
$1^{2}P_{3/2}$ & $\frac{3}{2}^{-}$ & 1871 & 1873 & 1823 \\
$1^{4}P_{1/2}$ & $\frac{1}{2}^{-}$ & 1894 & 1897 & \\
$1^{4}P_{3/2}$ & $\frac{3}{2}^{-}$ & 1879 & 1881 & 1823 \\
$1^{4}P_{5/2}$ & $\frac{5}{2}^{-}$ & 1859 & 1859 & \\
\hline
$2^{2}P_{1/2}$ & $\frac{1}{2}^{-}$ & 2361 & 2373 &  \\
$2^{2}P_{3/2}$ & $\frac{3}{2}^{-}$ & 2337 & 2347 & \\
$2^{4}P_{1/2}$ & $\frac{1}{2}^{-}$ & 2373 & 2386 &  \\
$2^{4}P_{3/2}$ & $\frac{3}{2}^{-}$ & 2349 & 2360 &  \\
$2^{4}P_{5/2}$ & $\frac{5}{2}^{-}$ & 2318 & 2325 &  \\
\hline
$3^{2}P_{1/2}$ & $\frac{1}{2}^{-}$ & 2929 & 2948 &  \\
$3^{2}P_{3/2}$ & $\frac{3}{2}^{-}$ & 2894 & 2913 & \\
$3^{4}P_{1/2}$ & $\frac{1}{2}^{-}$ & 2946 & 2966 &  \\
$3^{4}P_{3/2}$ & $\frac{3}{2}^{-}$ & 2912 & 2931 & \\
$3^{4}P_{5/2}$ & $\frac{5}{2}^{-}$ & 2865 & 2884 &  \\
\hline
$4^{2}P_{1/2}$ & $\frac{1}{2}^{-}$ & 3577 & 3609 &  \\
$4^{2}P_{3/2}$ & $\frac{3}{2}^{-}$ & 3532 & 3563 &  \\
$4^{4}P_{1/2}$ & $\frac{1}{2}^{-}$ & 3599 & 3632 & \\
$4^{4}P_{3/2}$ & $\frac{3}{2}^{-}$ & 3554 & 3586 & \\
$4^{4}P_{5/2}$ & $\frac{5}{2}^{-}$ & 3494 & 3524 & \\
\hline
\end{tabular}
\end{table*}

\begin{table*}
\centering
\caption{D-wave of $\Xi$ baryon (in MeV)}
\label{tab:d-wave}
\begin{tabular}{ccccc}
\hline
State & $J^{P}$ & $Mass_{cal}$1 & $Mass_{cal}$2 & $Mass_{exp}$\cite{pdg} \\
\hline
$1^{2}D_{3/2}$ & $\frac{3}{2}^{+}$ & 2270 & 2281 &   \\
$1^{2}D_{5/2}$ & $\frac{5}{2}^{+}$ & 2234 & 2244 & \\
$1^{4}D_{1/2}$ & $\frac{1}{2}^{+}$ & 2310 & 2322 & \\
$1^{4}D_{3/2}$ & $\frac{3}{2}^{+}$ & 2283 & 2295 &  \\
$1^{4}D_{5/2}$ & $\frac{5}{2}^{+}$ & 2247 & 2257 & \\
$1^{4}D_{7/2}$ & $\frac{7}{2}^{+}$ & 2203 & 2211 &  \\
\hline
$2^{2}D_{3/2}$ & $\frac{3}{2}^{+}$ & 2819 & 2842 &  \\
$2^{2}D_{5/2}$ & $\frac{5}{2}^{+}$ & 2771 & 2791 & \\
$2^{4}D_{1/2}$ & $\frac{1}{2}^{+}$ & 2874 & 2899 & \\
$2^{4}D_{3/2}$ & $\frac{3}{2}^{+}$ & 2838 & 2861 &  \\
$2^{4}D_{5/2}$ & $\frac{5}{2}^{+}$ & 2790 & 2810 & \\
$2^{4}D_{7/2}$ & $\frac{7}{2}^{+}$ & 2729 & 2747 & \\
\hline
$3^{2}D_{3/2}$ & $\frac{3}{2}^{+}$ & 3455 & 3489 &  \\
$3^{2}D_{5/2}$ & $\frac{5}{2}^{+}$ & 3391 & 3423 & \\
$3^{4}D_{1/2}$ & $\frac{1}{2}^{+}$ & 3527 & 3562 & \\
$3^{4}D_{3/2}$ & $\frac{3}{2}^{+}$ & 3479 & 3513 & \\
$3^{4}D_{5/2}$ & $\frac{5}{2}^{+}$ & 3415 & 3448 & \\
$3^{4}D_{7/2}$ & $\frac{7}{2}^{+}$ & 3336 & 3366 & \\
\hline
\end{tabular}
\end{table*}

\begin{table*}
\centering
\caption{F-wave of $\Xi$ baryon (in MeV)}
\label{tab:f-wave}
\begin{tabular}{ccccc}
\hline
State & $J^{P}$ & $Mass_{cal}$1 & $Mass_{cal}$2 & $Mass_{exp}$\cite{pdg}  \\
\hline
$1^{2}F_{5/2}$ & $\frac{5}{2}^{-}$ & 2713 & 2736 &  \\
$1^{2}F_{7/2}$ & $\frac{7}{2}^{-}$ & 2647 & 2666 &  \\
$1^{4}F_{3/2}$ & $\frac{3}{2}^{-}$ & 2786 & 2813 &  \\
$1^{4}F_{5/2}$ & $\frac{5}{2}^{-}$ & 2733 & 2757 & \\
$1^{4}F_{7/2}$ & $\frac{7}{2}^{-}$ & 2667 & 2687 &  \\
$1^{4}F_{9/2}$ & $\frac{9}{2}^{-}$ & 2588 & 2603 &  \\
\hline
$2^{2}F_{5/2}$ & $\frac{5}{2}^{-}$ & 3333 & 3368 & \\
$2^{2}F_{7/2}$ & $\frac{7}{2}^{-}$ & 3249 & 3280 & \\
$2^{4}F_{3/2}$ & $\frac{3}{2}^{-}$ & 3426 & 3465 & \\
$2^{4}F_{5/2}$ & $\frac{5}{2}^{-}$ & 3358 & 3394 & \\
$2^{4}F_{7/2}$ & $\frac{7}{2}^{-}$ & 3274 & 3306 & \\
$2^{4}F_{9/2}$ & $\frac{9}{2}^{-}$ & 3173 & 3201 & \\
\hline
\end{tabular}
\end{table*}

\begin{table*}
\centering
\caption{Comparison of masses with other predictions based on $J^{P}$ value for $\Xi$ baryon (in MeV)}
\label{tab:positive}
\begin{tabular}{ccccccccccccc}
\hline
$J^{P}$ & $Mass_{cal}1$ & $Mass_{cal}2$ & \cite{faustov} & \cite{loring} & \cite{capstick} & \cite{bijker} & \cite{melde} & \cite{santopinto} & \cite{oh} & \cite{pervin} & \cite{chen} & \cite{bgr}\\
\hline
$\frac{1}{2}^{+}$ & 1322 & 1321 & 1330 & 1310 & 1305 & 1334 & 1348 & 1317 & 1318 & 1325 & 1317 & 1303 $\pm$ 13 \\
 & 1884 & 1891 & 1886 & 1876 & 1840 & 1727 & 1805 & 1772 & 1932 & 1891 & 1750 & 2178 $\pm$ 48  \\
  & 2310 & 2322 & 1993 & 2062 & 2040 & 1932 & 1868 & & & & 1980 & 2231 $\pm$ 44 \\
  & 2361 & 2372 & 2012 & 2131 & 2100 & & & 1874 & & & 2054 & 2408 $\pm$ 45 \\
  & 2874 & 2899 & 2091 & 2176 & 2130 & & & & & & 2107 &  \\
  & 2935 & 2954  & 2142 & 2215 & 2150 & & & & & & 2149 & \\
  & 3527 & 3562 & 2367 & 2249 & 2230 & & & & & & 2254 &  \\
  & 3591 & 3620 & & & 2345 & & & & & & &  \\
\hline
$\frac{3}{2}^{+}$ & 1531 &  1524 & 1518 & 1539 & 1505 & 1524 & 1528 & 1552 & 1539 & 1520  & 1526 & 1553 $\pm$ 18 \\
 & 1971 & 1964 & 1966  & 1988 & 2045 & 1878 & & 1653 & 2120 & 1934 & 1952 & 2228 $\pm$ 40 \\
 & 2270 & 2281 & 2100 & 2076 & 2065 & 1979 & & & & & 1970 & 2398 $\pm$ 52 \\
 & 2283 & 2295 & 2121 & 2128 & 2115 & & & & & & 2065 & 2574 $\pm$ 52 \\
 & 2457 & 2459 & 2122 & 2170 & 2165 & & & & & & 2114 & \\
 & 2819 & 2842  & 2144 & 2175 & 2170 & & & & & & 2174 & \\
 & 2838 & 2861 & 2149 & 2219 & 2210 & & & & & & 2184 & \\ 
 & 3029 & 3041 & 2421  & 2257 & 2230 & & & & & & 2218 & \\
 & 3455 & 3489  & & 2279 & 2275 & & & & & & 2252 & \\
 & 3479 & 3513 \\
 & 3679 & 3702 \\
\hline
$\frac{5}{2}^{+}$ & 2234 & 2242 & 2108 & 2013 & 2045 & & & & & 1936 & 1959 & \\
 & 2247 & 2295 & 2147 & 2141 & 2165 & & & & & 2025 & 2102 & \\
 & 2771 & 2791 & 2213 & 2197 & 2230 & & & & & & 2170 &  \\
 & 2790 & 2810 & & 2231 & 2230 & & & & & & 2205 &  \\
 & 3391 & 3423 & & 2279 & 2240 & & & & & & 2239 & \\
 & 3415 & 3448 \\
\hline
$\frac{7}{2}^{+}$ & 2203 & 2211 & 2189 & 2169 & 2180 & & & & & 2035 & 2074 & \\
 & 2729 & 2747 & & 2289 & 2240 & & & & & & 2189 & \\
 & 3336 & 3366 \\
 \hline
\end{tabular}
\end{table*}

\begin{table*}
\centering
\caption{Comparison of masses with other predictions based on $J^{P}$ value $\Xi$ baryon (in MeV)}
\label{tab:negative}
\begin{tabular}{cccccccccccccc}
\hline
$J^{P}$ & $Mass_{cal}1$ & $Mass_{cal}2$ & \cite{faustov} & \cite{loring} & \cite{capstick} & \cite{bijker} & \cite{melde} & \cite{santopinto} & \cite{oh} & \cite{pervin} & \cite{chen} & \cite{bgr}\\
\hline
$\frac{1}{2}^{-}$ & 1886 & 1889 & 1682 & 1770 & 1755 & 1869 & & & 1658 & 1725 & 1772 & 1716 $\pm$ 43 \\
 & 1894 & 1897  & 1758 & 1922 & 1810 & 1932 & & & & 1811 & 1894 & 1837 $\pm$ 28 \\
 & 2361 & 2373 & 1839 & 1938 & 1835 & 2076 & & & & & 1926 & 1844 $\pm$ 43 \\
 & 2373 & 2386 & 2160 & 2241 & 2225  & & & & & & & 2758 $\pm$ 78 \\
 & 2929 & 2948 & 2210 & 2266 & 2285 & & & & & & & \\
 & 2946 & 2966 & 2233 & 2387 & 2300 & & & & & & &\\
 & 3577 & 3609 & 2261 & 2411 & 2320 & & & & & & & \\
 & 3599 & 3632  & & 2445 & 2380  & & & & & & & &\\
\hline 
$\frac{3}{2}^{-}$ & 1871 & 1873 & 1764 & 1780 & 1785 & 1828 & 1792 & 1861 & 1820 & 1759 & 1801 & 1906 $\pm$ 29 \\
 & 1879 & 1881 & 1798 & 1873 & 1880 & 1869 & & 1971 & & 1826 & 1918 & 1894 $\pm$ 38 \\
 & 2337 & 2347  & 1904 & 1924 & 1895 & 1932 & & & & & 1976 & 2497 $\pm$ 61 \\
 & 2349 & 2360 & 2245 & 2246 & 2240 & & & & & & & 2426 $\pm$ 73 \\
 & 2786 & 2813 & 2252 & 2284 & 2305 & & & & & & & \\
 & 2894 & 2913 & 2350 & 2353 & 2330& & & & & & & \\
 & 2912 & 2931 & 2352 & 2384 & 2340& & & & & & & \\
 & 3426 & 3465 & & 2416 & 2385 & & & & & & & \\
 & 3532 & 3563 & & & & & & & & & & \\
 & 3554 & 3586 & & & & & & & & & & \\
\hline
$\frac{5}{2}^{-}$ & 1859 & 1859 & 1853 & 1955 & 1900 & 1881 & & & & 1883 & 1917 &  \\
 & 2318 & 2325 & 2333 & 2292 &  2345 \\
 & 2713 & 2736 & 2411 & 2409 & 2350 \\
 & 2733 & 2757  & & 2425 & 2385 \\
 & 2865 & 2884 & & 2438 & \\
 & 3333 & 3368 \\
 & 3358 & 3394 \\
 & 3494 & 3524 \\
\hline
$\frac{7}{2}^{-}$ & 2647 & 2666 & 2460 & 2320 & 2355 \\
 & 2667 & 2687  & 2474 & & 2425 &  \\
 & 3249 & 3280 & &  2464 \\
 & 3274 & 3306 & & 2481 &  \\
\hline
$\frac{9}{2}^{-}$ & 2588 &  2603  & 2502 & 2505 & \\
 & 3173 & 3201  & & 2570 & \\
\hline
\end{tabular}
\end{table*}

An attempt has been made to summarize the calculated masses in present article along with those by different models. The following tables \ref{tab:positive}, \ref{tab:positive1}, \ref{tab:positive2}, \ref{tab:negative}, \ref{tab:negative1} and \ref{tab:negative2} depict the range of masses for a given $J^{P}$ value irrespective of assigned state in increasing order.  It is evident that the low lying resonance states are within considerable range for almost all the models and approaches listed. However, the higher states have huge variations possibly because of the fact that not a single model exactly predicts the spin-parity assignments and in addition to it there are no experimental evidence for the states. Also, the present calculations have included masses upto 4 GeV. \\\\
The Ref. \cite{faustov} has employed relativistic quark-diquark model for the calculation of strange baryon mass spectra. As the model considers both ground and excited states of diquarks, the number of excited states are limited and only confined to lower states. The another relativistic approach based on three quark Bethe-Salpeter equation with instantaneous two and three body forces is described by Ref. \cite{loring}. It has introduced instanton induced hyperfine splitting of positive and negative parity states. Another approach is well known relativised Capstick-Isgur model with higher order spin-dependent potential terms in three quark system and has predicted masses for nearly 2 GeV \cite{capstick}.\\\\
The approach by R.Bijker {\it et al.} \cite{bijker} uses the algebraic model. It is based on collective string-like qqq wherein the orbital excitations are treated as rotations and vibrations of the strings. Low-lying states are established by this model for octet and decuplet class but exact spin-parity are not assigned in the case of $\Xi$. Ref. \cite{melde} has utilised relativistic constituent quark model (RCQM) with Goldstone-boson exchange. Another relativistic quark-diquark approach with Coulomb plus linear interaction along with an exchange term which is inspired by G\"{u}rsey-Radicati has been employed in Ref. \cite{santopinto} for low-lying resonance states of $\Xi$. Y. Oh \cite{oh} has studied the cascade and omega baryons through Skyrme model which is based on the approximation of equal mass splitting of hyperon resonances. The mass formula is developed with isospin and spin in the soliton-kaon bound-state model. The $\Xi$ has also been explored in large-$N_{c}$ limit \cite{matagne,goity} as well as through QCD-SUM Rule method \cite{lee}. M. Pervin \cite{pervin} has obtained mass spectra using a non-relativistic quark model approach. Y. Chen et al. has implemented a different approach with non-relativistic quark model supplemented with hyperfine interaction of higher order $O(\alpha_{s}^{2})$ \cite{chen}. Some low-lying states have been exploited through dynamical chirally improved quarks by BGR Collaboration \cite{bgr}.  \\\\
The ground state of $\Xi$ has been very well established with known spin-parity at 1321 MeV and $J^{P}=\frac{1}{2}^{+}$ in the octet family. It is evident from table 5 that ground state fits well for nearly all the models owing the little variations to the assumptions of any given model. Another state is 1532 MeV with $J^{P}=\frac{3}{2}^{+}$ holding a place in decuplet. The mass for this state varies within 20 MeV among all the models discussed. The only negative parity state by PDG is 1823 MeV at $J^{P}=\frac{3}{2}^{-}$ which is obtained as 1871 MeV and 1879 MeV for octet and decuplet $\Xi$.\\\\
$\Xi$(1690) is fairly known state in the PDG database however the spin-parity assignments and exact mass predictions vary a lot. The BABAR Collaboration concluded the state to be spin $\frac{1}{2}$ \cite{babar}. As shown in Table 6, various models have predicted this state in a comparatively higher mass from PDG the nearest being 1682 MeV. Also, due to intrinsic drawback of the present model, it could not provide conclusive assignment of this state. The $\Xi$(2030) with assigned angular momentum value to be $\frac{5}{2}$ is predicted here to be 2234 MeV with positive parity. Other models predictions vary within 200 MeV range for the same spin-parity. $\Xi$(1620) appears in PDG with one star status. However, any such state is not established in this work but one study throws the light on the existence of $\Xi$(1620) and $\Xi$(1690) \cite{ramos}. \\\\
The PDG states $\Xi$(1950), $\Xi$(2250) and $\Xi$(2370) are three and two starred however due to lack of spin-parity assignment, the comparison is not reasonable. In addition, one study depicts the states of cascade around $\Xi$(1950) within the range of 1900-2000 MeV into three different states \cite{pavon}. Few of the states of our results are comparatively near to the BGR work \cite{bgr} which don't appear in other approaches. 

\begin{table*}
\centering
\caption{S-wave of $\Lambda$ baryon (in MeV)}
\begin{tabular}{ccccc}
\hline
State & $J^{P}$ & $Mass_{cal}$1 & $Mass_{cal}$2 & $Mass_{exp}$\cite{pdg} \\
\hline
1S & $\frac{1}{2}^{+}$ & 1115 & 1115 & 1115 \\
2S & $\frac{1}{2}^{+}$ & 1592 & 1589 & 1600 \\
3S & $\frac{1}{2}^{+}$ & 1885 & 1892 & 1810 \\
4S & $\frac{1}{2}^{+}$ & 2202 & 2220 & \\
5S & $\frac{1}{2}^{+}$ & 2540 &2571 & \\
\hline
\end{tabular}
\caption{P-wave of $\Lambda$ baryon (in MeV)}
\begin{tabular}{ccccc}
\hline
State & $J^{P}$ & $Mass_{cal}$1 & $Mass_{cal}$2 & $Mass_{exp}$ \cite{pdg}\\
\hline
$1^{2}P_{1/2}$ & $\frac{1}{2}^{-}$ & 1546 & 1558 & 1670\\
$1^{2}P_{3/2}$ & $\frac{3}{2}^{-}$ & 1534 & 1544 & 1520\\
$1^{4}P_{1/2}$ & $\frac{1}{2}^{-}$ & 1553 & 1564 & \\
$1^{4}P_{3/2}$ & $\frac{3}{2}^{-}$ & 1540 & 1551 & \\
$1^{4}P_{5/2}$ & $\frac{5}{2}^{-}$ & 1524 & 1533 & \\
\hline
$2^{2}P_{1/2}$ & $\frac{1}{2}^{-}$ & 1834 & 1858 & 1800 \\
$2^{2}P_{3/2}$ & $\frac{3}{2}^{-}$ & 1819 & 1841 & 1690 \\
$2^{4}P_{1/2}$ & $\frac{1}{2}^{-}$ & 1841 & 1867 & \\
$2^{4}P_{3/2}$ & $\frac{3}{2}^{-}$ & 1827 & 1850 & \\
$2^{4}P_{5/2}$ & $\frac{5}{2}^{-}$ & 1807 & 1827 & 1830 \\
\hline
$3^{2}P_{1/2}$ & $\frac{1}{2}^{-}$ & 2149 & 2186 & \\
$3^{2}P_{3/2}$ & $\frac{3}{2}^{-}$ & 2131 & 2166 & \\
$3^{4}P_{1/2}$ & $\frac{1}{2}^{-}$ & 2158 & 2196 & \\
$3^{4}P_{3/2}$ & $\frac{3}{2}^{-}$ & 2140 & 2176 & \\
$3^{4}P_{5/2}$ & $\frac{5}{2}^{-}$ & 2116 & 2149 & \\
\hline
$4^{2}P_{1/2}$ & $\frac{1}{2}^{-}$ & 2484 & 2536 & \\
$4^{2}P_{3/2}$ & $\frac{3}{2}^{-}$ & 2464 & 2513 & \\
$4^{4}P_{1/2}$ & $\frac{1}{2}^{-}$ & 2495 & 2548 & \\
$4^{4}P_{3/2}$ & $\frac{3}{2}^{-}$ & 2474 & 2525 & \\
$4^{4}P_{5/2}$ & $\frac{5}{2}^{-}$ & 2447 & 2494 & \\
\hline
\end{tabular}
\end{table*}

\begin{table*}
\centering
\caption{D-wave of $\Lambda$ baryon (in MeV)}
\begin{tabular}{ccccc}
\hline
State & $J^{P}$ & $Mass_{cal}$1 & $Mass_{cal}$2 & $Mass_{exp}$\cite{pdg} \\
\hline
$1^{2}D_{3/2}$ & $\frac{3}{2}^{+}$ & 1769 & 1789 & 1890\\
$1^{2}D_{5/2}$ & $\frac{5}{2}^{+}$ & 1746 & 1767 & 1820\\
$1^{4}D_{1/2}$ & $\frac{1}{2}^{+}$ & 1794 & 1814 & \\
$1^{4}D_{3/2}$ & $\frac{3}{2}^{+}$ & 1777 & 1798 & \\
$1^{4}D_{5/2}$ & $\frac{5}{2}^{+}$ & 1755 & 1776 & \\
$1^{4}D_{7/2}$ & $\frac{7}{2}^{+}$ & 1727 & 1748 & \\
\hline
$2^{2}D_{3/2}$ & $\frac{3}{2}^{+}$ & 2076 & 2113 & 2070 \\
$2^{2}D_{5/2}$ & $\frac{5}{2}^{+}$ & 2051 & 2085 & 2110 \\
$2^{4}D_{1/2}$ & $\frac{1}{2}^{+}$ & 2105 & 2144 & \\
$2^{4}D_{3/2}$ & $\frac{3}{2}^{+}$ & 2086 & 2123 & \\
$2^{4}D_{5/2}$ & $\frac{5}{2}^{+}$ & 2060 & 2096 & \\
$2^{4}D_{7/2}$ & $\frac{7}{2}^{+}$ & 2029 & 2061 & 2085 \\
\hline
$3^{2}D_{3/2}$ & $\frac{3}{2}^{+}$ & 2408 & 2459 & \\
$3^{2}D_{5/2}$ & $\frac{5}{2}^{+}$ & 2378 & 2426 & \\
$3^{4}D_{1/2}$ & $\frac{1}{2}^{+}$ & 2441 & 2496 & \\
$3^{4}D_{3/2}$ & $\frac{3}{2}^{+}$ & 2419 & 2471 & \\
$3^{4}D_{5/2}$ & $\frac{5}{2}^{+}$ & 2389 & 2438 & \\
$3^{4}D_{7/2}$ & $\frac{7}{2}^{+}$ & 2352 & 2398 & \\
\hline
\end{tabular}
\caption{F-wave of $\Lambda$ baryon (in MeV)}
\begin{tabular}{ccccc}
\hline
State & $J^{P}$ & $Mass_{cal}$1 & $Mass_{cal}$2 & $Mass_{exp}$\cite{pdg}  \\
\hline
$1^{2}F_{5/2}$ & $\frac{5}{2}^{-}$ & 2005 & 2039 & \\
$1^{2}F_{7/2}$ & $\frac{7}{2}^{-}$ & 1970 & 2002 & 2100 \\
$1^{4}F_{3/2}$ & $\frac{3}{2}^{-}$ & 2043 & 2079 & \\
$1^{4}F_{5/2}$ & $\frac{5}{2}^{-}$ & 2015 & 2050 & \\
$1^{4}F_{7/2}$ & $\frac{7}{2}^{-}$ & 1980 & 2013 & \\
$1^{4}F_{9/2}$ & $\frac{9}{2}^{-}$ & 1939 & 1969 & \\
\hline
$2^{2}F_{5/2}$ & $\frac{5}{2}^{-}$ & 2329 & 2380 & \\
$2^{2}F_{7/2}$ & $\frac{7}{2}^{-}$ & 2291 & 2337 & \\
$2^{4}F_{3/2}$ & $\frac{3}{2}^{-}$ & 2371 & 2427 & \\
$2^{4}F_{5/2}$ & $\frac{5}{2}^{-}$ & 2341 & 2393 & \\
$2^{4}F_{7/2}$ & $\frac{7}{2}^{-}$ & 2303 & 2350 & \\
$2^{4}F_{9/2}$ & $\frac{9}{2}^{-}$ & 2257 & 2299 & \\
\hline
$3^{2}F_{5/2}$ & $\frac{5}{2}^{-}$ & 2676 & 2741 & \\
$3^{2}F_{7/2}$ & $\frac{7}{2}^{-}$ & 2632 & 2693 & \\
$3^{4}F_{3/2}$ & $\frac{3}{2}^{-}$ & 2723 & 2793 & \\
$3^{4}F_{5/2}$ & $\frac{5}{2}^{-}$ & 2689 & 2755 & \\
$3^{4}F_{7/2}$ & $\frac{7}{2}^{-}$ & 2645 & 2707 & \\
$3^{4}F_{9/2}$ & $\frac{9}{2}^{-}$ & 2593 & 2650 & \\
\hline
\end{tabular}
\end{table*}

\begin{table*}
\caption{G-wave of $\Lambda$ baryon (in MeV)}
\centering
\begin{tabular}{ccccc}
\hline
State & $J^{P}$ & $Mass_{cal}$1 & $Mass_{cal}$2 & $Mass_{exp}$\cite{pdg}  \\
\hline
$1^{2}G_{7/2}$ & $\frac{7}{2}^{+}$ & 2253 & 2302 & \\
$1^{2}G_{9/2}$ & $\frac{9}{2}^{+}$ & 2204 & 2246 & 2350 \\
$1^{4}G_{5/2}$ & $\frac{5}{2}^{+}$ & 2305 & 2363 & \\
$1^{4}G_{7/2}$ & $\frac{7}{2}^{+}$ & 2265 & 2316 & \\
$1^{4}G_{9/2}$ & $\frac{9}{2}^{+}$ & 2216 & 2260 & \\
$1^{4}G_{11/2}$ & $\frac{11}{2}^{+}$ & 2159 & 2195 & \\
\hline
\end{tabular}
\end{table*}

\begin{table*}
%\centering
\caption{Comparison of masses with other predictions based on $J^{P}$ value for $\Lambda$ baryon (in MeV)}
\label{tab:positive1}
\begin{tabular}{cccccccccccccc}
\hline
$J^{P}$ & $Mass_{cal}1$ & $Mass_{cal}2$ & \cite{faustov} & \cite{loring} & \cite{capstick} & \cite{bijker} & \cite{melde} & \cite{santopinto} & \cite{chen1} & \cite{amiri} & \cite{chen} & \cite{bgr}\\
\hline
$\frac{1}{2}^{+}$ & 1115 & 1115 & 1115 & 1108 & 1115 & 1133 & 1136 & 1116 & 1116 & 1112 & 1113 & 1149 $\pm$ 18 \\
 & 1592 & 1589 & 1615 & 1677 & 1680 & 1577 & 1625 & 1518 & 1600 & 1695 & 1606 & 1807 $\pm$ 94 \\
 & 1794 & 1814 & & & & 1799 & 1799 & 1666 & 1810 & & 1764 & 2112 $\pm$ 54 \\
 & 1885 & 1892 & 1901 & 1747 & 1830 & & & 1955 & & & 1880 & 2137 $\pm$ 68 \\
 & 2105 & 2144 & 1972 & 1898 & 1910 & & & 1960 & & & 2013 & \\
 & 2202 & 2220 & 1986 & 2077 & 2010 & & & & & & 2173 & \\
 & 2441 & 2496 & 2042 & 2099 & 2105 & & & & & & 2198 & \\
 & 2540 & 2571 & 2099 & 2132 & 2120 & & & & & & & \\
\hline
$\frac{3}{2}^{+}$ & 1769 & 1789 & 1854 & 1823 & 1900 & 1849 & & 1896 & 1890 & 1903 & 1836 & 1991 $\pm$ 103 \\
 & 1777 & 1798 & 1976 & 1952 & 1960 & & & & 2000 & & 1958 & 2058 $\pm$ 139 \\
 & 2076 & 2113 & 2130 & 2045 & 1995 & & & & & 1993 & & 2481 $\pm$ 111 \\
 & 2086 & 2123 & 2184 & 2087 & 2050 & & & & & 2061 & \\
 & 2408 & 2459 & 2202 & 2133 & 2080 & & & & & 2121 & \\
 & 2419 & 2471 & & & & & & & & & 2134 & \\
 \hline
$\frac{5}{2}^{+}$ & 1746 & 1767 & 1825 & 1834 & 1890 & 1849 & & 1896 & 1820 & 1846 & 1839 & \\
 & 1755 & 1776 & 2098 & 1999 & 2035 & 2074 & & & 2110 & 2132 & 2008 & \\
 & 2051 & 2085 & 2221 & 2078 & 2115 & & & & & & 2103 & \\
 & 2060 & 2096 & 2255 & 2127 & 2115 & & & & & & 2129 & \\
 & 2305 & 2363 & 2258 & 2150 & 2180 & & & & & & 2155 & \\
 & 2378 & 2426 & & & & & & & & & & \\
 & 2389 & 2471 & & & & & & & & & & \\
 \hline
$\frac{7}{2}^{+}$ & 1727 & 1748 & 2251 & 2130 & 2120 & & & & 2020 & & 2064 & \\
 & 2029 & 2061 & 2471 & 2331 & & & & & & & & \\
 & 2253 & 2302 & & & & & & & & & & \\
 & 2265 & 2316 & & & & & & & & & & \\
 & 2352 & 2398 & & & & & & & & & & \\
\hline
$\frac{9}{2}^{+}$ & 2204 & 2246 & 2360 & 2340 & & 2357 & & & 2350 & 2360 & & \\
 & 2216 & 2260 & & & & & & & & & & \\
\hline
$\frac{11}{2}^{+}$ & 2159 & 2195 & & & & & & & 2585 & & & \\
\hline
\end{tabular}
\end{table*}

\begin{table*}
\centering
\caption{Comparison of masses with other predictions based on $J^{P}$ value for $\Lambda$ baryon (in MeV)}
\label{tab:negative1}
\begin{tabular}{ccccccccccccccc}
\hline
$J^{P}$ & $Mass_{cal}1$ & $Mass_{cal}2$ & \cite{faustov} & \cite{loring} & \cite{capstick} & \cite{bijker} & \cite{melde} & \cite{santopinto} & \cite{chen1} & \cite{amiri} & \cite{chen} & \cite{bgr}\\
\hline
$\frac{1}{2}^{-}$ & 1546 & 1558 & 1406 & 1524 & 1550 & 1686 & 1556 & 1650 & 1670 & 1679 & 1559 & 1416 $\pm$ 81 \\
 & 1553 & 1564 & 1667 & 1630 & 1615 & 1799 & 1682 & 1732 & & 1830 & 1656 & 1546 $\pm$ 110 \\
 & 1834 & 1858 & 1733 & 1816 & 1675 & & 1778 & 1969 & 1800 & & 1791 & 1713 $\pm$ 116 \\
 & 1841 & 1867 & 1927 & 2011 & 2015 & & & 1854 & & & & 2075 $\pm$ 249 \\
 & 2149 & 2186 & 2197 & 2076 & 2095 & & & 1928 & & & & \\
 & 2158 & 2196 & 2218 & 2117 & 2160 & & & & & & & \\
 & 2484 & 2536 & & & & & & & & & & \\
 & 2495 & 2548 & & & & & & & & & & \\
\hline
$\frac{3}{2}^{-}$ & 1534 & 1544 & 1549 & 1508 & 1545 & 1686 & 1556 & 1650 & 1690 & 1683 & 1560 & 1751 $\pm$ 41 \\
 & 1540 & 1551 & 1693 & 1662 & 1645 & & 1682 & 1785 & 1520 & & 1702 & 2203 $\pm$ 106 \\
 & 1819 & 1841 & 1812 & 1775 & 1770 & & & 1854 & 2325 & & 1859 & 2381 $\pm$ 87 \\
 & 1827 & 1850 & 2035 & 1987 & 2030 & & & 1928 & & & & \\
 & 2043 & 2079 & 2319 & 2090 & 2110 & & & 1969 & & & & \\
 & 2131 & 2166 & 2322 & 2147 & 2185 & & & & & & & \\
 & 2140 & 2176 & 2392 & 2259 & 2230 & & & & & & & \\
 & 2371 & 2427 & 2454 & 2290 & 2275 & & & & & & & \\
 & 2464 & 2513 & 2468 & 2313 & & & & & & & & \\
 & 2474 & 2525 & & & & & & & & & & \\
 & 2723 & 2793 & & & & & & & & & & \\
\hline
$\frac{5}{2}^{-}$ & 1524 & 1533 & & & & & & & &  & & & \\
 & 1807 & 1827 & 1861 & 1828 & 1775 & 1799 & 1778 & 1785 & 1830 & 1850 & 1803 & \\
 & 2005 & 2039 & 2136 & 2080 & 2180 & & & & & & & \\
 & 2015 & 2050 & 2350 & 2179 & 2250 & & & & & & & \\
 & 2116 & 2149 & & & & & & & & & & \\
 & 2329 & 2380 & & & & & & & & & & \\
 & 2341 & 2393 & & & & & & & & & & \\
 & 2447 & 2494 & & & & & & & & & & \\
 & 2676 & 2741 & & & & & & & & & & \\
 & 2689 & 2755 & & & & & & & & & & \\
\hline 
$\frac{7}{2}^{-}$ & 1970 & 2002 & 2097 & 2090 & 2150 & & & & 2100 & 2087 & & \\
 & 1980 & 2013 & 2583 & 2227 & 2230 & & & & & & & \\
 & 2291 & 2337 & & & & & & & & & & \\
 & 2303 & 2350 & & & & & & & & & & \\
 & 2632 & 2693 & & & & & & & & & & \\
 & 2645 & 2707 & & & & & & & & & & \\
\hline 
$\frac{9}{2}^{-}$ & 1939 & 1969 & 2665 & 2370 & & & & & & & \\
 & 2257 & 2299 & & & & & & & & & & \\
 & 2593 & 2650 & & & & & & & & & & \\
\hline
\end{tabular}
\end{table*}  

\begin{table*}
\centering
\caption{S-wave of $\Sigma$ baryon (in MeV)}
\begin{tabular}{ccccc}
\hline
State & $J^{P}$ & $Mass_{cal}$1 &  $Mass_{cal}$2 & $Mass_{exp}$\cite{pdg} \\
\hline
1S & $\frac{1}{2}^{+}$ & 1193 & 1193 & 1193 \\
 & $\frac{3}{2}^{+}$ & 1384 &  1384 & 1385 \\
2S & $\frac{1}{2}^{+}$ & 1643 & 1643 &  1660 \\
  & $\frac{3}{2}^{+}$ & 1827 & 1827 & \\
3S & $\frac{1}{2}^{+}$ & 2083 & 2099 & \\
 & $\frac{3}{2}^{+}$ & 2229 & 2236 & \\
4S & $\frac{1}{2}^{+}$ & 2560 & 2589 & \\
 & $\frac{3}{2}^{+}$ & 2675 &  2693 & \\ 
5S & $\frac{1}{2}^{+}$ & 3067 & 3108 & \\
 & $\frac{3}{2}^{+}$ & 3159 & 3189 & \\
\hline
\end{tabular}
\caption{P-wave $\Sigma$ baryon (in MeV)}
\begin{tabular}{ccccc}
\hline
State & $J^{P}$ & $Mass_{cal}$1 &$Mass_{cal}$2 & $Mass_{exp}$\cite{pdg} \\
\hline
$1^{2}P_{1/2}$ & $\frac{1}{2}^{-}$ & 1720 & 1725 & 1620 \\
$1^{2}P_{3/2}$ & $\frac{3}{2}^{-}$ & 1698 & 1702 & 1670\\
$1^{4}P_{1/2}$ & $\frac{1}{2}^{-}$ & 1731 & 1736 & 1750 \\
$1^{4}P_{3/2}$ & $\frac{3}{2}^{-}$ & 1709 & 1713 & \\
$1^{4}P_{5/2}$ & $\frac{5}{2}^{-}$ & 1680 & 1683 & 1775 \\
\hline
$2^{2}P_{1/2}$ & $\frac{1}{2}^{-}$ & 2128 & 2145 &  1900 \\
$2^{2}P_{3/2}$ & $\frac{3}{2}^{-}$ & 2099 & 2114 & 1910 \\
$2^{4}P_{1/2}$ & $\frac{1}{2}^{-}$ & 2142 & 2159 & \\
$2^{4}P_{3/2}$ & $\frac{3}{2}^{-}$ & 2114 & 2129 & \\
$2^{4}P_{5/2}$ & $\frac{5}{2}^{-}$ & 2076 & 2087 & \\
\hline
$3^{2}P_{1/2}$ & $\frac{1}{2}^{-}$ & 2580 & 2608 & \\
$3^{2}P_{3/2}$ & $\frac{3}{2}^{-}$ & 2545 & 2571 & \\
$3^{4}P_{1/2}$ & $\frac{1}{2}^{-}$ & 2598 & 2627 & \\
$3^{4}P_{3/2}$ & $\frac{3}{2}^{-}$ & 2563 & 2589 & \\
$3^{4}P_{5/2}$ & $\frac{5}{2}^{-}$ & 2516 & 2541 & \\
\hline
$4^{2}P_{1/2}$ & $\frac{1}{2}^{-}$ & 3068 & 3111 & \\
$4^{2}P_{3/2}$ & $\frac{3}{2}^{-}$ & 3027 & 3067 & \\
$4^{4}P_{1/2}$ & $\frac{1}{2}^{-}$ & 3088 & 3133 & \\
$4^{4}P_{3/2}$ & $\frac{3}{2}^{-}$ & 3047 & 3089 & \\
$4^{4}P_{5/2}$ & $\frac{5}{2}^{-}$ & 2994 & 3030 & \\
\hline
$5^{2}P_{1/2}$ & $\frac{1}{2}^{-}$ & 3588 & 3641 & \\
$5^{2}P_{3/2}$ & $\frac{3}{2}^{-}$ & 3541 & 3594 & \\
$5^{4}P_{1/2}$ & $\frac{1}{2}^{-}$ & 3612 & 3664 & \\
$5^{4}P_{3/2}$ & $\frac{3}{2}^{-}$ & 3565 & 3617 & \\
$5^{4}P_{5/2}$ & $\frac{5}{2}^{-}$ & 3502 & 3555 & \\
\hline
\end{tabular}
\end{table*}

\begin{table*}
\centering
\caption{D-wave $\Sigma$ baryon (in MeV)}
\begin{tabular}{ccccc}
\hline
State & $J^{P}$ & $Mass_{cal}$1 & $Mass_{cal}$2 & $Mass_{exp}$\cite{pdg} \\
\hline
$1^{2}D_{3/2}$ & $\frac{3}{2}^{+}$ & 2040 & 2057 & 1940 \\
$1^{2}D_{5/2}$ & $\frac{5}{2}^{+}$ & 1998 & 2013 & 1915 \\
$1^{4}D_{1/2}$ & $\frac{1}{2}^{+}$ & 2086 & 2107 & \\
$1^{4}D_{3/2}$ & $\frac{3}{2}^{+}$ & 2055 & 2074 & \\
$1^{4}D_{5/2}$ & $\frac{5}{2}^{+}$ & 2014 & 2029 & \\
$1^{4}D_{7/2}$ & $\frac{7}{2}^{+}$ & 1962 & 1974 & 2025 \\
\hline
$2^{2}D_{3/2}$ & $\frac{3}{2}^{+}$ & 2481 & 2510 & \\
$2^{2}D_{5/2}$ & $\frac{5}{2}^{+}$ & 2432 & 2459 & \\
$2^{4}D_{1/2}$ & $\frac{1}{2}^{+}$ & 2536 & 2568 & \\
$2^{4}D_{3/2}$ & $\frac{3}{2}^{+}$ & 2499 & 2529 & \\
$2^{4}D_{5/2}$ & $\frac{5}{2}^{+}$ & 2451 & 2478 & \\
$2^{4}D_{7/2}$ & $\frac{7}{2}^{+}$ & 2390 & 2414 & \\
\hline
$3^{2}D_{3/2}$ & $\frac{3}{2}^{+}$ & 2962 & 3004 & \\
$3^{2}D_{5/2}$ & $\frac{5}{2}^{+}$ & 2905 & 2945 & \\
$3^{4}D_{1/2}$ & $\frac{1}{2}^{+}$ & 3027 & 3072 & \\
$3^{4}D_{3/2}$ & $\frac{3}{2}^{+}$ & 2984 & 3027 & \\
$3^{4}D_{5/2}$ & $\frac{5}{2}^{+}$ & 2926 & 2967 & \\
$3^{4}D_{7/2}$ & $\frac{7}{2}^{+}$ & 2855 & 2892 & \\
\hline
$4^{2}D_{3/2}$ & $\frac{3}{2}^{+}$ & 3476 & 3534 & \\
$4^{2}D_{5/2}$ & $\frac{5}{2}^{+}$ & 3410 & 3464 & \\
$4^{4}D_{1/2}$ & $\frac{1}{2}^{+}$ & 3549 & 3613 & \\
$4^{4}D_{3/2}$ & $\frac{3}{2}^{+}$ & 3500 & 3560 & \\
$4^{4}D_{5/2}$ & $\frac{5}{2}^{+}$ & 3435 & 3490 & \\
$4^{4}D_{7/2}$ & $\frac{7}{2}^{+}$ & 3353 & 3403 & \\
\hline
\end{tabular}
\caption{F-wave $\Sigma$ baryon (in MeV)}
\begin{tabular}{ccccc}
\hline
State & $J^{P}$ & $Mass_{cal}$1 & $Mass_{cal}$2 & $Mass_{exp}$\cite{pdg}  \\
\hline
$1^{2}F_{5/2}$ & $\frac{5}{2}^{-}$ & 2386 & 2416 & \\
$1^{2}F_{7/2}$ & $\frac{7}{2}^{-}$ & 2318 & 2343 & \\
$1^{4}F_{3/2}$ & $\frac{3}{2}^{-}$ & 2461 & 2495 & \\
$1^{4}F_{5/2}$ & $\frac{5}{2}^{-}$ & 2406 & 2437 & \\
$1^{4}F_{7/2}$ & $\frac{7}{2}^{-}$ & 2338 & 2365 & \\
$1^{4}F_{9/2}$ & $\frac{9}{2}^{-}$ & 2257 & 2278 & \\
\hline
$2^{2}F_{5/2}$ & $\frac{5}{2}^{-}$ & 2858 & 2901 & \\
$2^{2}F_{7/2}$ & $\frac{7}{2}^{-}$ & 2781 & 2819 & \\
$2^{4}F_{3/2}$ & $\frac{3}{2}^{-}$ & 2943 & 2990 & \\
$2^{4}F_{5/2}$ & $\frac{5}{2}^{-}$ & 2881 & 2925 & \\
$2^{4}F_{7/2}$ & $\frac{7}{2}^{-}$ & 2804 & 2844 & \\
$2^{4}F_{9/2}$ & $\frac{9}{2}^{-}$ & 2712 & 2746 & \\
\hline
$3^{2}F_{5/2}$ & $\frac{5}{2}^{-}$ & 3363 & 3417 & \\
$3^{2}F_{7/2}$ & $\frac{7}{2}^{-}$ & 3278 & 3330 & \\
$3^{4}F_{3/2}$ & $\frac{3}{2}^{-}$ & 3456 & 3512 & \\
$3^{4}F_{5/2}$ & $\frac{5}{2}^{-}$ & 3388 & 3443 & \\
$3^{4}F_{7/2}$ & $\frac{7}{2}^{-}$ & 3303 & 3356 & \\
$3^{4}F_{9/2}$ & $\frac{9}{2}^{-}$ & 3202 & 3252 & \\
\hline
\end{tabular}
\end{table*}

\begin{table*}
\centering
\caption{Comparison of masses with other predictions based on $J^{P}$ value for $\Sigma$ baryon (in MeV)}
\label{tab:positive2}
\begin{tabular}{cccccccccccccc}
\hline
$J^{P}$ & $Mass_{cal}1$ & $Mass_{cal}2$ & \cite{faustov} & \cite{loring} & \cite{capstick} & \cite{bijker} & \cite{melde} & \cite{santopinto} & \cite{chen1} & \cite{amiri} & \cite{chen} & \cite{bgr}\\
\hline
$\frac{1}{2}^{+}$ & 1193 & 1193 & 1187 & 1190 & 1190 & 1170 & 1180 & 1211 & 1193 & 1198 & 1192 & 1216 $\pm$ 15 \\
 & 1643 & 1643 & 1711 & 1760 & 1720 & 1604 & 1616 & 1546 & 1660 & 1656 & 1664 & 2069 $\pm$ 74 \\
 & 2083 & 2099 & 1922 & 1947 & 1915 &  & 1911 & 1668 & 1770 & & 1924 & 2149 $\pm$ 66\\
 & 2086 & 2107 & 1983 & 2009 & 1970 & & & 1801 & 1880 & & 1986 & 2335 $\pm$ 63 \\
 & 2536 & 2568 & 2028 & 2052 & 2005 & & & & & & 2022 & & \\
 & 2560 & 2589 & 2180 & 2098 & 2030 & & & & & & 2069 & & \\
 & 3027 & 3072 & 2292 & 2138 & 2105 & & & & & & 2172 & & \\
 & 3067 & 3108 & 2472 &  & & & & & & & & & \\
 & 3549 & 3613 & & & & & & & & & & & \\
\hline
$\frac{3}{2}^{+}$ & 1384 & 1384 & 1381 & 1411 & 1370 & 1382 & 1389 & 1334 & 1385 & 1381 &  1383 & 1471 $\pm$ 23 \\
 & 1827 & 1827 & 1862 & 1896 & 1920 & & 1865 & 1439 & 1560 & & 1868 & 2194 $\pm$ 81 \\
 & 2040 & 2057 & 2025 & 1961 & 1970 & & & 1924 & 1690 & & 1947 & 2250 $\pm$ 79 \\
 & 2055 & 2074 & 2076 & 2011 & 2010 & & & & 1840 & & 1993 & 2468 $\pm$ 67 \\
 & 2229 & 2236 & 2096 & 2044 & 2030 & & & & & & 2039 & & \\
 & 2481 & 2510 & 2157 & 2062 & 2045 & & & & & & 2075 & & \\
 & 2499 & 2529 & 2186 & 2103 & 2085 & & & & & & 2098 & & \\
 & 2675 & 2693 & & 2112 & 2115 & & & & & & 2122 & & \\
 & 2962 & 3004 & & & & & & & & & 2168 & & \\
 & 2984 & 3027 & & & & & & & & & & & \\
 & 3159 & 3189 & & & & & & & & & & & \\
\hline
$\frac{5}{2}^{+}$ & 1998 & 2013 & 1991 & 1956 & 1995 & 1872 & & 2061 & 1915 & 1930 & 1949 & & \\
 & 2014 & 2029 & 2062 & 2027 & 2030 & & & & 2070 & & 2028 & & \\
 & 2432 & 2459 & 2221 & 2071 & 2095 & & & & & & 2062 & & \\
 & 2451 & 2478 & & & & & & & & & 2107 & & \\
 & 2905 & 2945 & & & & & & & & & 2154 & & \\
 & 2926 & 2967 & & & & & & & & & & & \\
 & 3410 & 3464 & & & & & & & & & & & \\
 & 3435 & 3490 & & & & & & & & & & & \\
\hline
$\frac{7}{2}^{+}$ & 1962 & 1974 & 2033 & 2070 & 2060 & 2012 & & & 2030 & 2039 & 2002 & & \\
 & 2390 & 2414 & 2470 & 2161 & 2125 & & & & & & 2106 & & \\
 & 2855 & 2892 & & & & & & & & & & & \\
 & 3353 & 3403 & & & & & & & & & & & \\
 \hline
\end{tabular}
\end{table*}

\begin{table*}
\centering
\caption{Comparison of masses with other predictions based on $J^{P}$ value for $\Sigma$ baryon (in MeV)}
\label{tab:negative2}
\begin{tabular}{ccccccccccccccc}
\hline
$J^{P}$ & $Mass_{cal}1$ & $Mass_{cal}2$ & \cite{faustov} & \cite{loring} & \cite{capstick} & \cite{bijker} & \cite{melde} & \cite{santopinto} & \cite{chen1} & \cite{amiri} & \cite{chen} & \cite{bgr}\\
\hline
$\frac{1}{2}^{-}$ & 1720 & 1725 & 1620 & 1628 & 1630 & 1711 & 1677 & 1753 & 1620 & 1754 & 1657 & 1603 $\pm$ 38 \\
 & 1731 & 1736 & 1693 & 1771 & 1675 & & 1736 & 1868 & 1750 & & 1746 & 1718 $\pm$ 58 \\
 & 2128 & 2145 & 1747 & 1798 & 1695 & & 1759 & 1895 & 2000 & & 1802 & 1730 $\pm$ 34 \\
 & 2142 & 2159 & 2115 & 2111 & 2110 & & & & & & & 2478 $\pm$ 104 \\
 & 2580 & 2608 & 2198 & 2136 & 2155 & & & & & & & & \\
 & 2598 & 2627 & 2202 & 2251 & 2165 & & & & & & & & \\
 & 3068 & 3111 & 2289 & 2264 & 2205 & & & & & & & & \\
 & 3088 & 3133 & 2381 & 2288 & 2260 & & & & & & & & \\
 & 3588 & 3641 & & & & & & & & & & & \\
 & 3612 & 3664 & & & & & & & & & & & \\
\hline 
$\frac{3}{2}^{-}$ & 1698 & 1702 & 1706 & 1669 & 1655 & 1711 & 1677 & 1753 & 1670 & 1697 & 1698 &  1861 $\pm$ 26 \\
 & 1709 & 1713 & 1731 & 1728 & 1750 & 1974 & 1736 & 1868 & 1940 & 1956 & 1790 & 1736 $\pm$ 40 \\
 & 2099 & 2114 & 1856 & 1781 & 1755 & & 1759 & 1895 & 2250 & & 1802 & 2394 $\pm$ 74 \\
 & 2114 & 2129 & 2175 & 2139 & 2120 & & & & & & & 2297 $\pm$ 122 \\
 & 2416 & 2495 & 2203 & 2171 & 2185 & & & & & & & & \\
 & 2545 & 2571 & 2300 & 2203 & 2200 & & & & & & & & \\
 & 2563 & 2589 & & & & & & & & & & & \\
 & 2943 & 2990 & & & & & & & & & & & \\
 & 3027 & 3067 & & & & & & & & & & & \\
 & 3047 & 3089 & & & & & & & & & & & \\
 & 3456 & 3512 & & & & & & & & & & & \\
 & 3541 & 3594 & & & & & & & & & & & \\
 & 3565 & 3617 & & & & & & & & & & & \\
\hline
$\frac{5}{2}^{-}$ & 1680 & 1683 & 1757 & 1770 & 1755 & & 1736 & 1753 & 1775 & 1777 & 1743 & & \\
 & 2076 & 2087 & 2214 & 2174 & 2205 & & & & & & & & \\
 & 2386 & 2416 & 2347 & 2226 & 2250 & & & & & & & & \\
 & 2406 & 2437 & & & & & & & & & & & \\
 & 2516 & 2541 & & & & & & & & & & & \\
 & 2858 & 2901 & & & & & & & & & & & \\
 & 2881 & 2925 & & & & & & & & & & & \\
 & 2994 & 3030 & & & & & & & & & & & \\
 & 3363 & 3417 & & & & & & & & & & & \\
 & 3388 & 3443 & & & & & & & & & & & \\
 & 3565 & 3617 & & & & & & & & & & & \\
\hline
$\frac{7}{2}^{-}$ & 2318 & 2343 & 2259 & 2236 & 2245 & & & & 2100 & & & & \\
 & 2338 & 2365 & 2349 & 2285 & & & & & & & & & \\
 & 2781 & 2819 & & & & & & & & & & & \\
 & 2804 & 2844 & & & & & & & & & & & \\
 & 3278 & 3330 & & & & & & & & & & & \\ 
 & 3303 & 3356 & & & & & & & & & & & \\
\hline
$\frac{9}{2}^{-}$ & 2257 & 2278 & 2289 & 2325 & & & & & & & & & \\
 & 2712 & 2746 & & & & & & & & & & & \\
 & 3202 & 3252 & & & & & & & & & & & \\
\hline
\end{tabular}
\end{table*}
{\bf 
For $\Lambda$ baryon, the ground state mass is 1115 MeV and for $\Sigma$ it is 1193 MeV and confinement parameters are determined accordingly. Here, the constituent quark mass for u and d quarks is similar so, the charges of $\Sigma$ are not distinguished.  The four star status states are in good agreement with the PDG masses as evident from the tables. As for excited states of $\Lambda$ 2S(1600), the predicted masses are very near to almost all the models' mass. However, the first negative parity state $\Lambda$(1405) $J^{P}=\frac{1}{2}^{-}$  is not established by the present model but the next state with $J^{P}=\frac{3}{2}^{-}$ 1520 MeV is varies by 15 MeV from PDG. The $J^{P}=\frac{5}{2}^{+}$ state of 1D(1769 MeV) too falls within the PDG mass range of 1750-1850. The $J^{P}=\frac{9}{2}^{+}$ for 1G is somewhat under-predicted owing to the limitations of the hCQM. \\\\
For the case of $\Sigma$ baryon, the early negative parity states are just one star status. The later state with $J^{P}=\frac{1}{2}^{-}$ appearing as $\Sigma$(1750) is predicted well here as 1720 MeV within the range 1700-1800. The $J^{P}=\frac{5}{2}^{+}$ $\Sigma$(1915) is slightly higher predicted from most of the models. The other higher ranged states results from hCQM are comparable to those of BGR Collaboration results \cite{bgr}.  }

\section{Regge Trajectories}
Regge trajectories have been one of the useful tools in spectroscopic studies. The plot of total angular momentum J and principle quantum number n against the square of resonance mass $M^{2}$ are drawn based on calculated data. The non-intersecting and linearly fitted lines have been in accordance with theoretical and experimental data in many studies \cite{z16}. These plots might be helpful in predicting the correct spin-parity assignment of a given state. 
\begin{subequations}
\begin{align}
J = aM^{2} + a_{0} \\
n = b M^{2} + b_{0}
\end{align}
\end{subequations}

\begin{figure*}
\centering
\caption{\label{fig:xi-nm} Regge trajectory $\Xi$ for $n \rightarrow M^{2}$}
\includegraphics[scale=0.5]{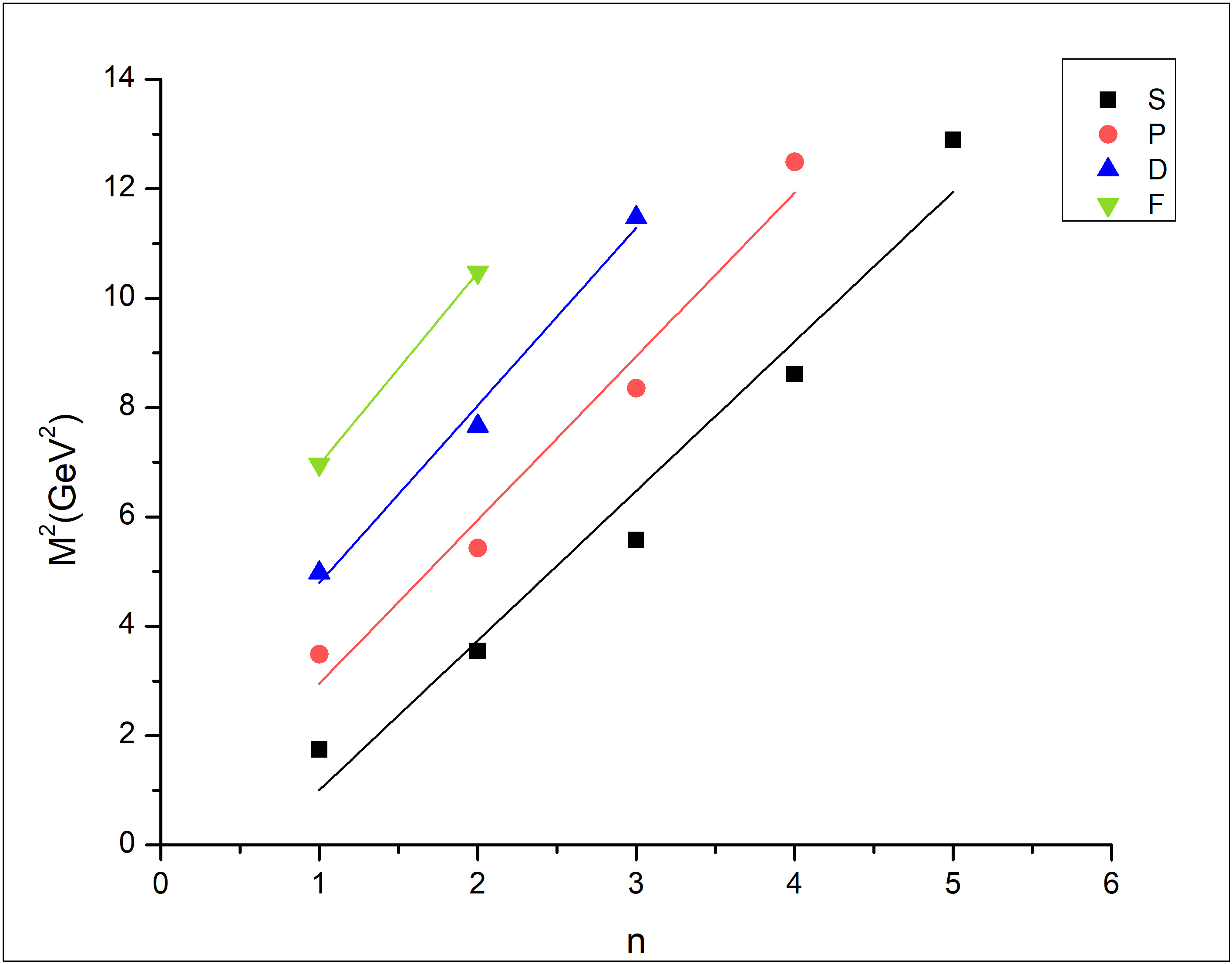}
\end{figure*}

\begin{figure*}
\centering
\caption{\label{fig:xi-jm1} Regge trajectory $\Xi$ for $J^{P} \rightarrow M^{2}$}
\includegraphics[scale=0.5]{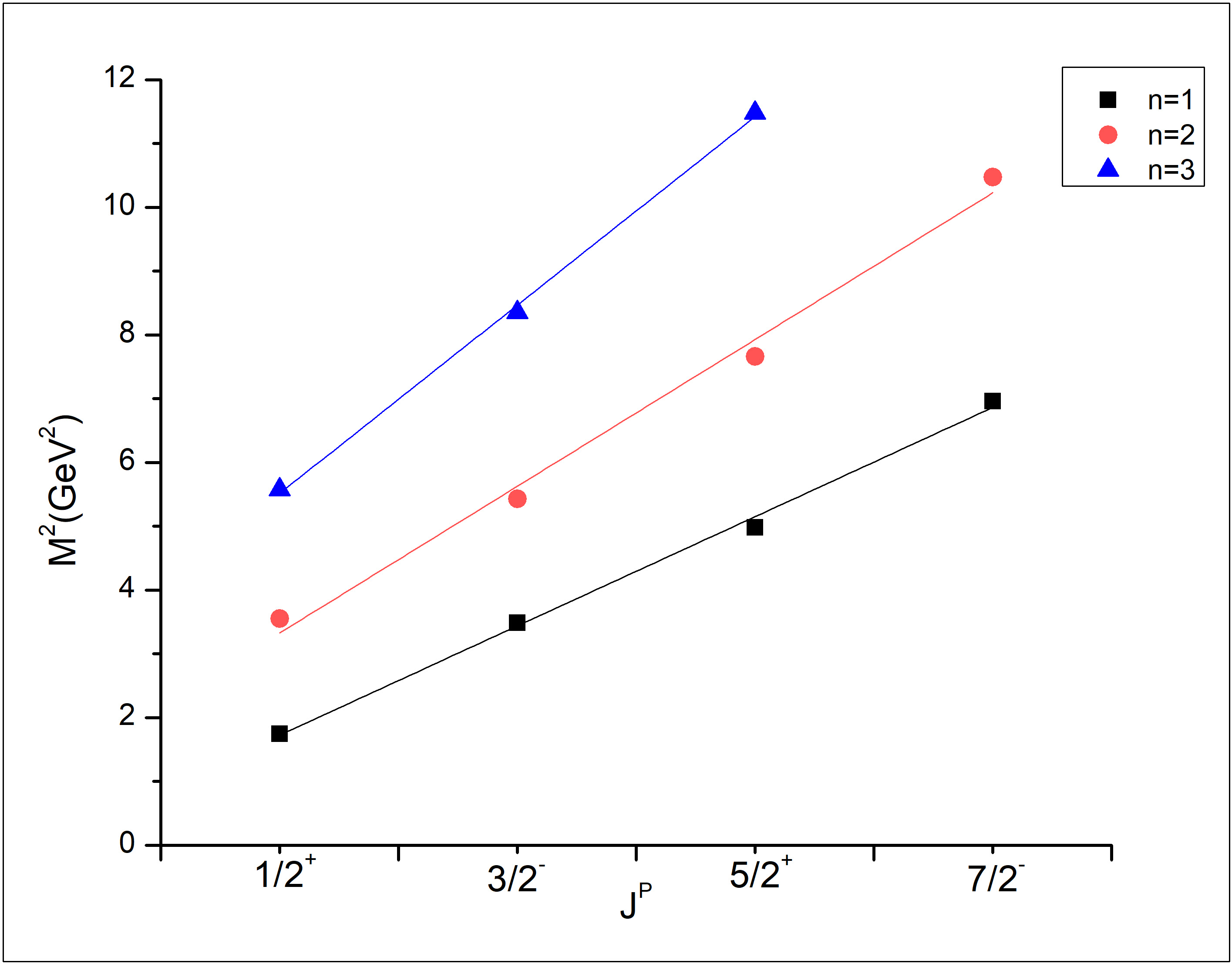}
\end{figure*}

\begin{figure*}
\centering
\caption{\label{fig:xi-jm2} Regge trajectory $\Xi$ for $J^{P} \rightarrow M^{2}$}
\includegraphics[scale=0.5]{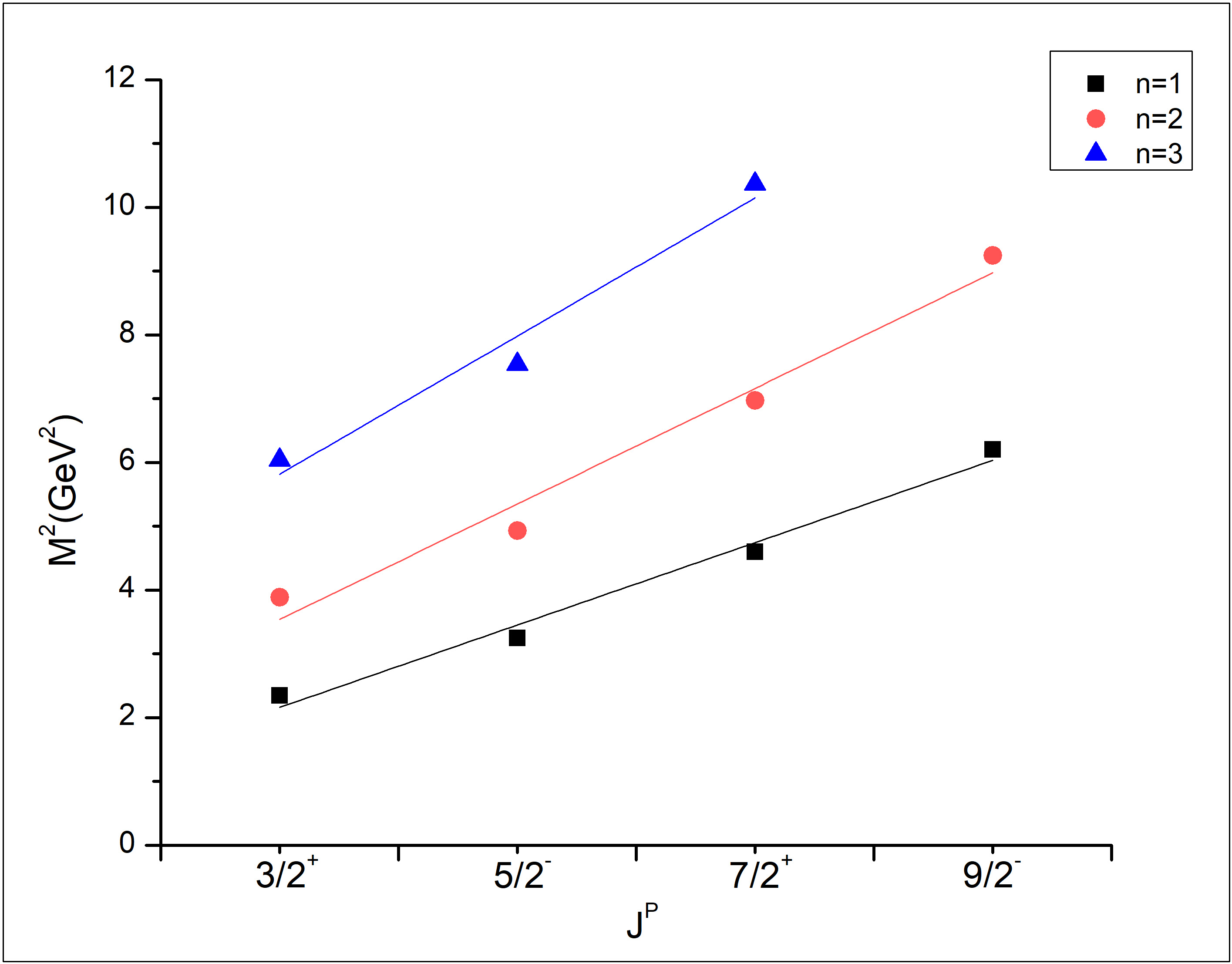}
\end{figure*}

\begin{figure*}
\centering
\caption{\label{fig:sigma-nm} Regge trajectory $\Sigma$ for $n \rightarrow M^{2}$}
\includegraphics[scale=0.5]{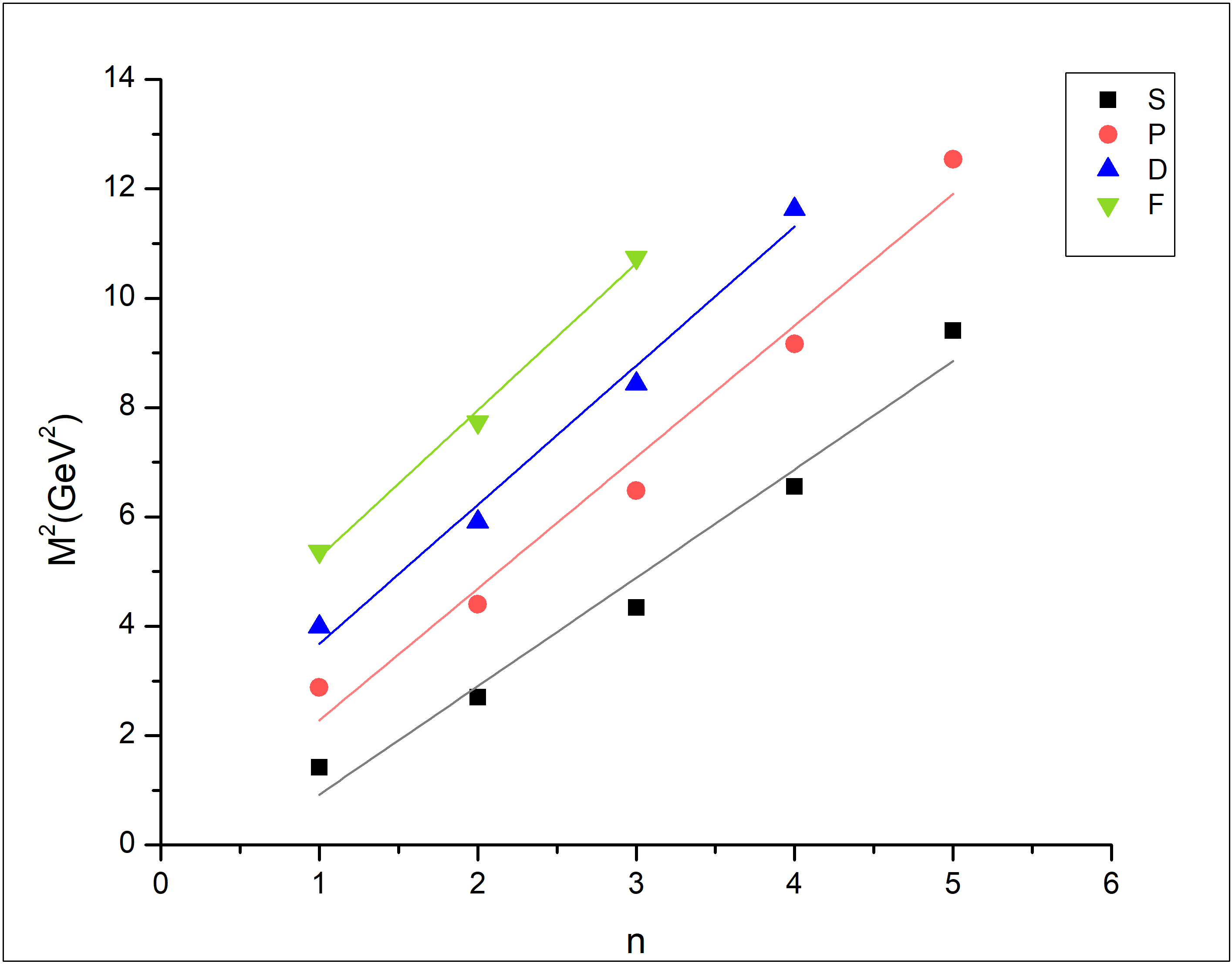}
\end{figure*}

\begin{figure*}
\centering
\caption{\label{fig:lambda-nm} Regge trajectory $\Lambda$ for $n \rightarrow M^{2}$}
\includegraphics[scale=0.5]{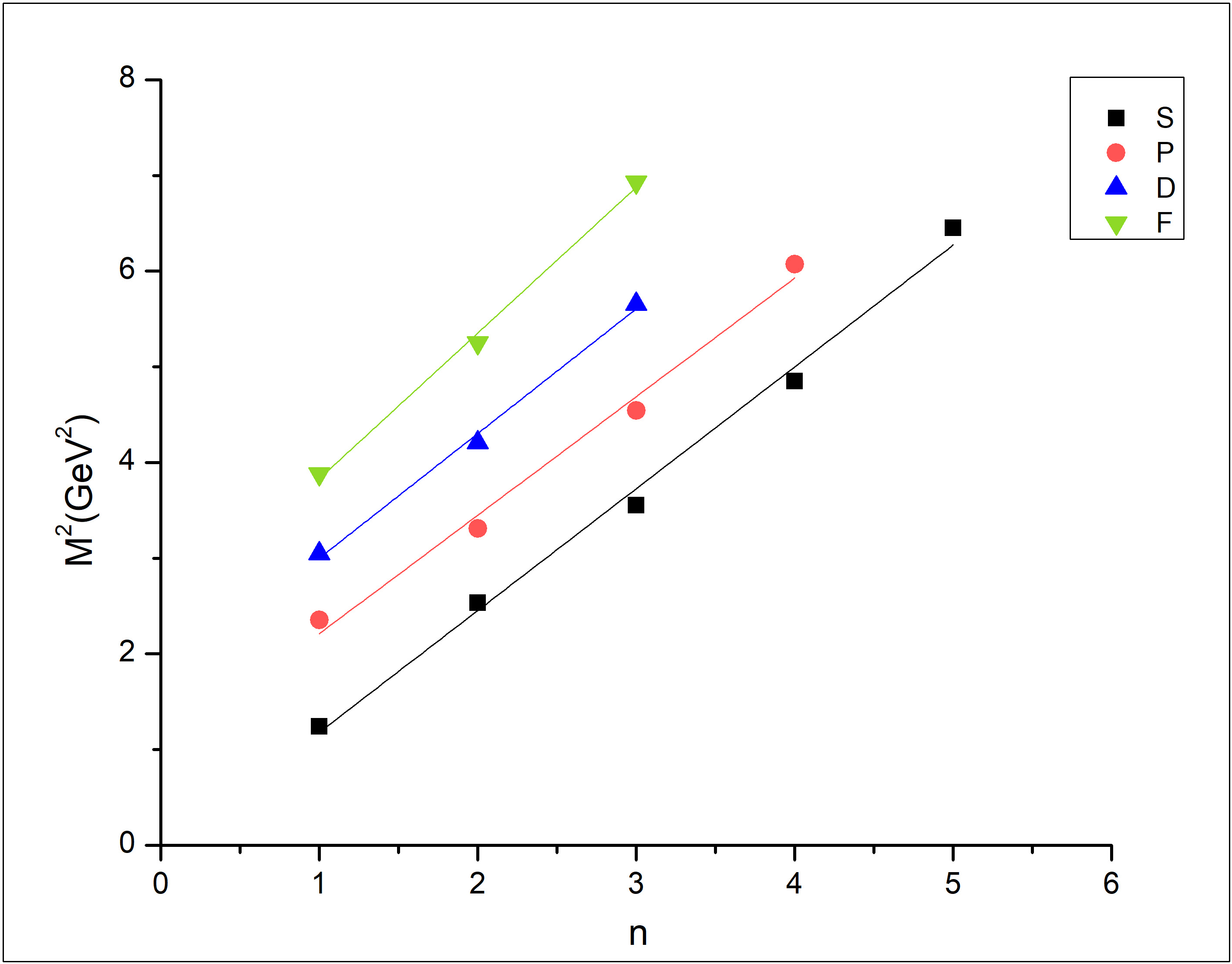}
\end{figure*}

As it is evident from the graphs {[\ref{fig:xi-nm}], [\ref{fig:sigma-nm}] and [\ref{fig:lambda-nm}}], the Regge trajectory for n against $M^{2}$ has been linearly fitted for the calculated resonance masses which follows the expected trend. The trajectories for J against $M^{2}$ for the natural parity also follows the linear nature which signifies that the spin-parity assignment for the obtained states are in agreement. The overall nature of the Regge trajectories observed for baryon studies agrees with the calculated results for few of the states. 

\begin{figure*}
\centering
\caption{\label{fig:sigma-12} Regge trajectory $\Sigma$ for $J^{P} \rightarrow M^{2}$}
\includegraphics[scale=0.5]{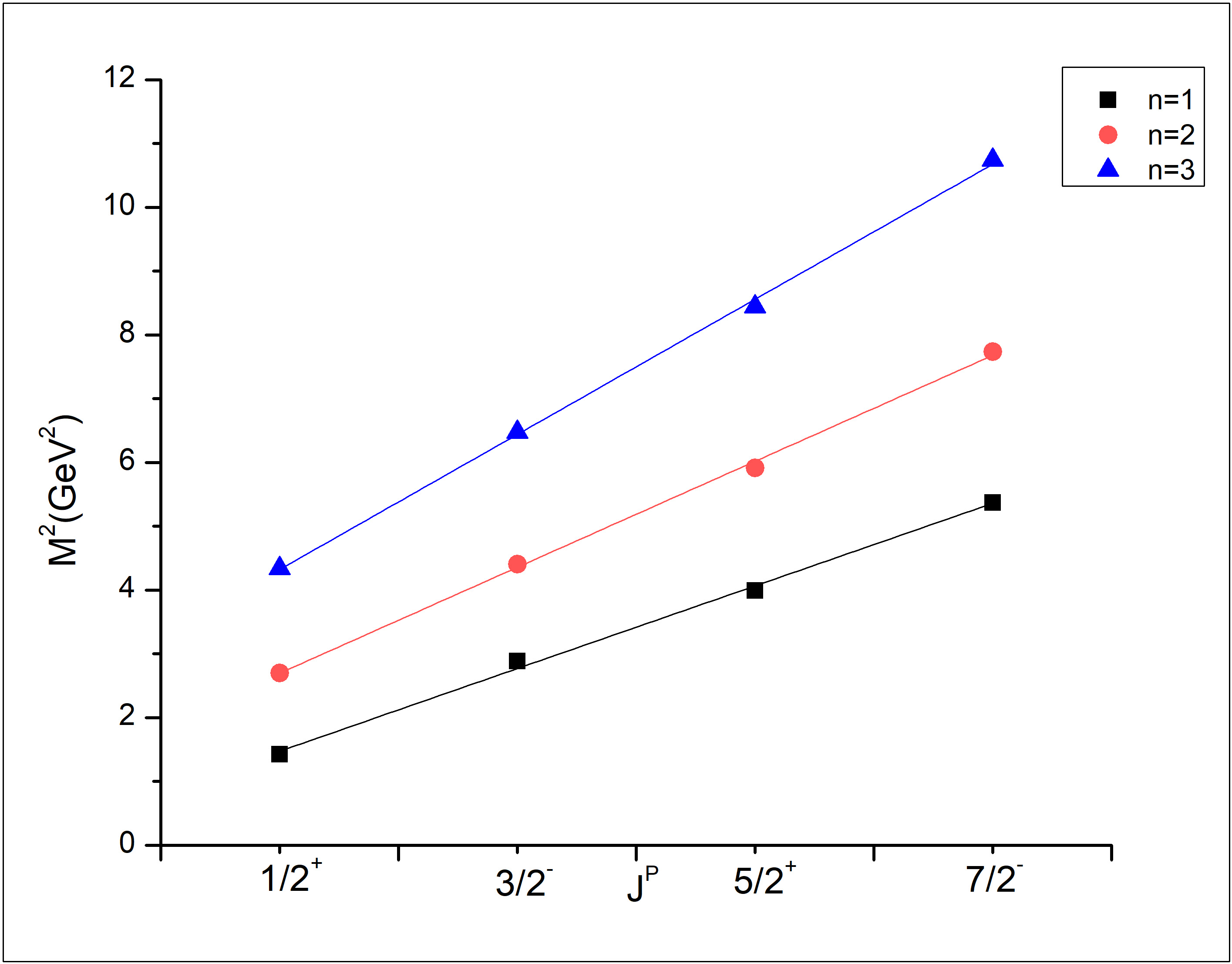}
\end{figure*}

\begin{figure*}
\centering
\caption{\label{fig:sigma-32} Regge trajectory $\Sigma$ for $J^{P} \rightarrow M^{2}$}
\includegraphics[scale=0.5]{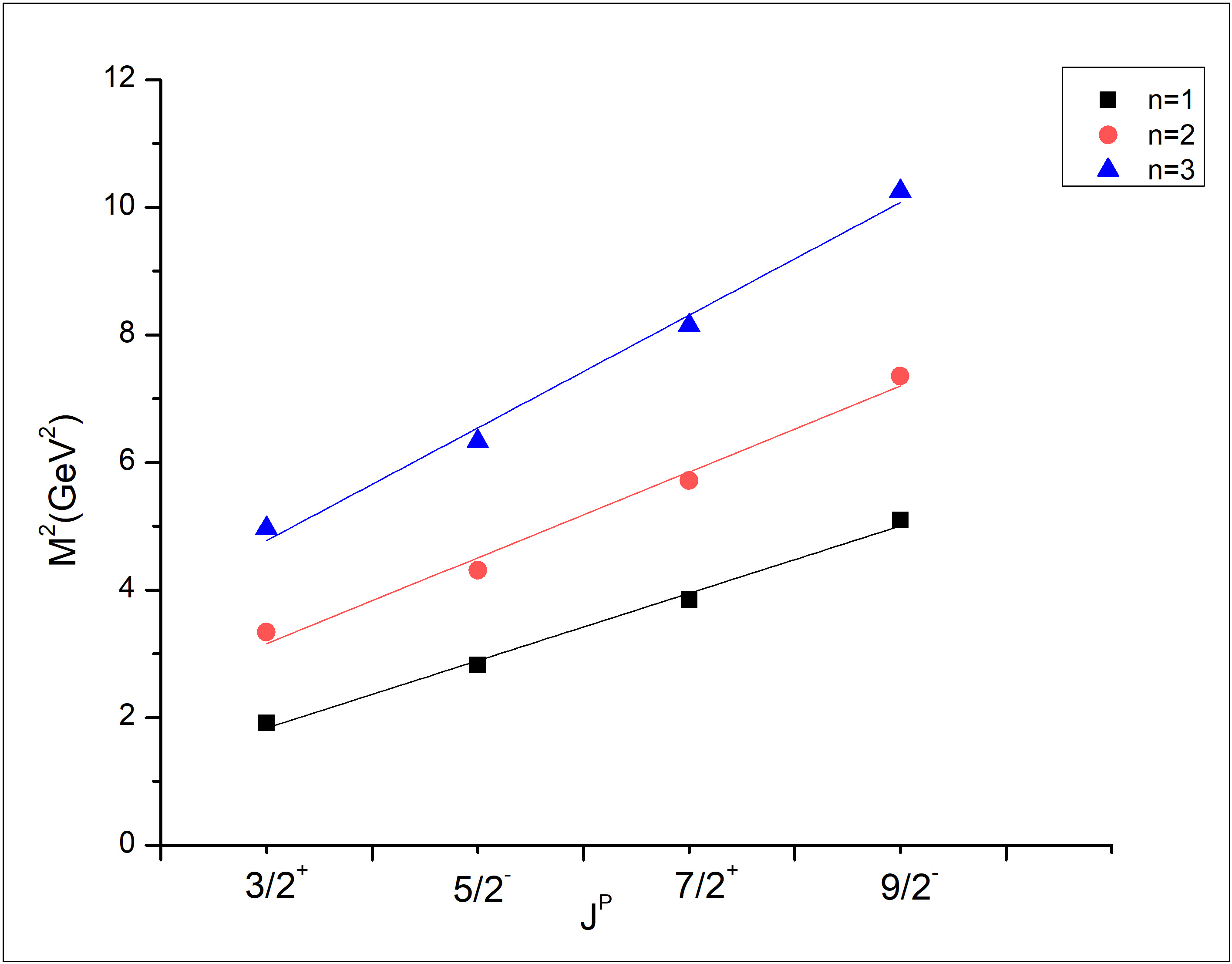}
\end{figure*}

\begin{figure*}
\centering
\caption{\label{fig:lambda-12} Regge trajectory $\Lambda$ for $J^{P} \rightarrow M^{2}$}
\includegraphics[scale=0.5]{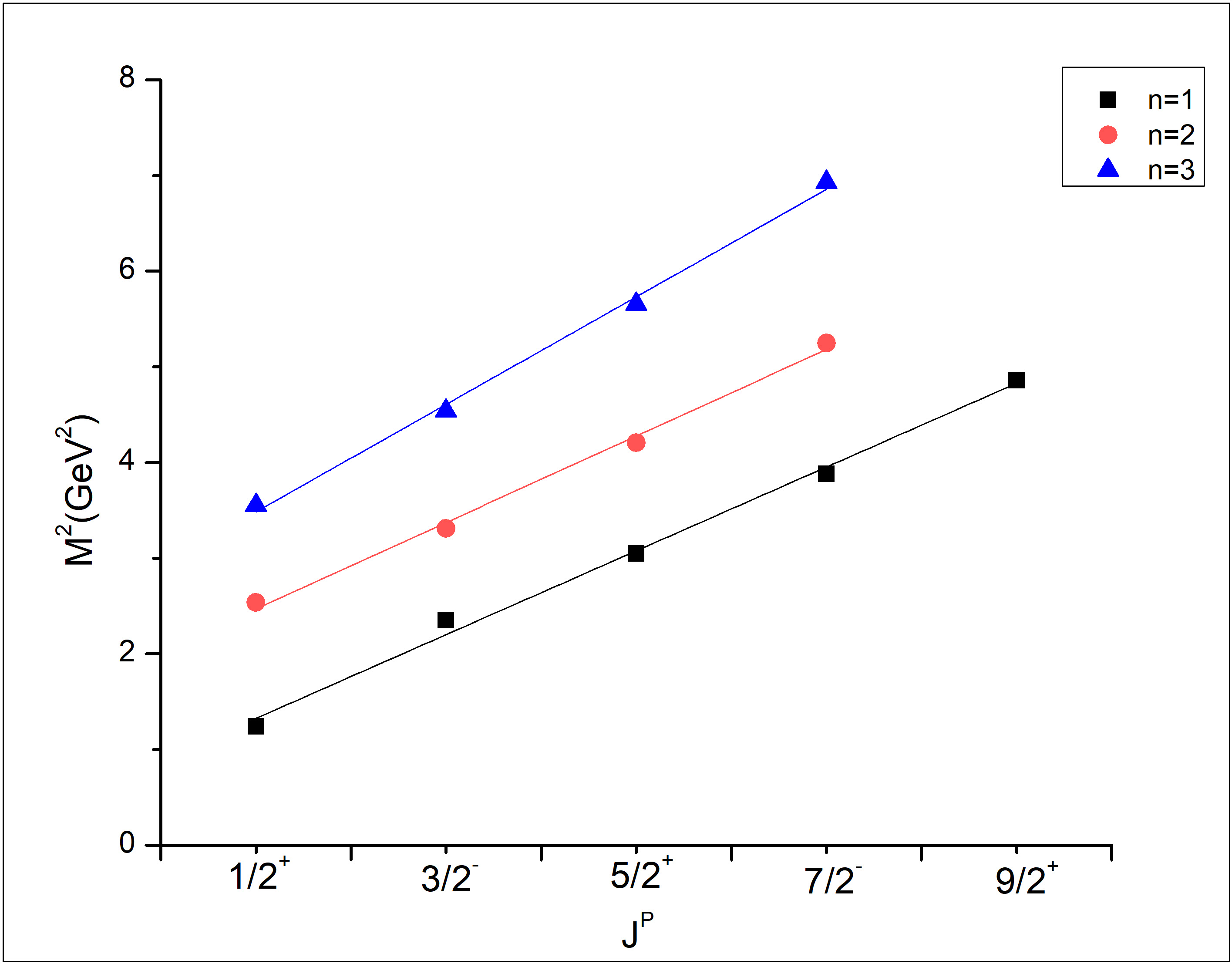}
\end{figure*}

\section{Magnetic Moment}
The study of electromagnetic properties of baryons is an active area for theoretical as well as experimental work. This intrinsic property help reveal the shape and other dynamics of transition in decay modes. In case of cascade $\Xi$ and $\Sigma$ baryon, magnetic moment is to be determined for both the spin configuration based on the respective spin-flavour wave-function. The generalized form of magentic moment is
\begin{equation}
\mu_{B}= \sum_{q} \left\langle \phi_{sf} \vert \mu_{qz}\vert \phi_{sf} \right\rangle
\end{equation}
where $\phi_{sf}$ is the spin-flavour wave function. The contribution from individual quark appears as 
\begin{equation}
\mu_{qz}= \frac{e_{q}}{2m_{q}^{eff}}\sigma_{qz}
\end{equation}
$e_{q}$ being the quark charge, $\sigma_{qz}$ being the spin orientation and $m_{q}^{eff}$ is the effective mass which may vary from model based quark mass due to interactions. Here, is it noteworthy that magnetic moment shall have contribution from many other effects within the baryon as sea quark, valence quark, orbital etc. Various models have contributed to obtaining the octet and decuplet baryons' magnetic moment. 
The order of quarks in spin-flavour won't be affecting the magnetic moment calculation. The final spin-flavour wave function along with calculated ground state magnetic moment in terms of nuclear magneton ($\mu_{N}$) are mentioned in the following table [\ref{tab:mm}]. \\

\begin{table*}
\centering
\caption{Magnetic moment for ground states}
\label{tab:mm}
\begin{tabular}{ccccc}
\hline
Spin & Baryon & $\sigma_{qz}$ & Mass (MeV) & $\mu$ ($\mu_{N}$)\\
\hline
$\frac{1}{2}$ & $\Xi^{0}$(uss) & $\frac{1}{3}(4\mu_{s}-\mu_{u})$ & 1322 & -1.50 \\
$\frac{1}{2}$ & $\Xi^{-}$(dss) & $\frac{1}{3}(4\mu_{s}-\mu_{d})$ & 1322 & -0.46\\
$\frac{3}{2}$ & $\Xi^{*0}$(uss) & $(2\mu_{s}+\mu_{u})$ & 1531 & 0.766 \\
$\frac{3}{2}$ & $\Xi^{*-}$(dss) & $(2\mu_{s}+\mu_{d})$ & 1531 & -1.962 \\
$\frac{1}{2}$ & $\Sigma^{+}$(uus)& $\frac{1}{3}(4\mu_{u}-\mu_{s})$ & 1193 & 2.79\\
$\frac{1}{2}$ & $\Sigma^{0}$(uds)& $\frac{1}{3}(2\mu_{u}+2\mu_{d}-\mu_{s})$ & 1193 & 0.839 \\
$\frac{1}{2}$ & $\Sigma^{-}$(dds)& $\frac{1}{3}(4\mu_{d}-\mu_{s})$ & 1193 & -1.113 \\
$\frac{3}{2}$ & $\Sigma^{*+}$(uus)& $(2\mu_{u}+\mu_{s})$ & 1384 & 2.877 \\
$\frac{3}{2}$ & $\Sigma^{*0}$(uds)& $(\mu_{u}+\mu_{d}+\mu_{s})$ & 1384 & 0.353 \\
$\frac{3}{2}$ & $\Sigma^{*-}$(dds)& $(2\mu_{d}+\mu_{s})$ & 1384 & -2.171 \\
$\frac{1}{2}$ & $\Lambda^{0}$(uds)& $\mu_{s}$ & 1115 & -0.606 \\
\hline
\end{tabular}
\end{table*}

\begin{table*}
\centering
\caption{Comparison of calculated magnetic moments with various models (All data in units of $\mu_{N}$)}
\label{tab:mmcompare}
\begin{tabular}{ccccccccccccc}
\hline
Baryon & $\mu_{cal}$ & Exp \cite{pdg} & \cite{harleen3} & \cite{harleen3} & \cite{linde} & \cite{aliev} & \cite{dhir} & \cite{dhir} & \cite{hong} & \cite{flee} & \cite{harpreet} & \cite{buchman} \\
\hline
$\Xi^{0}$ & -1.50 & -1.25 & -1.41 & -1.39 & & -1.3 & & & -1.25  & -1.37 & & \\
$\Xi^{-}$ & -0.46 & -0.651 & -0.50 & -0.50 & & -0.7 & &  & -1.07 & -0.82 & & \\
$\Xi^{*0}$ & 0.766 & & 0.60 & 0.49 &  0.49 & 0.69 & 0.48 & 0.32 & 0.44 & 0.16 & 0.508 & 0.65 \\
$\Xi^{*-}$ & -1.962 & & -2.11 & -2.43 &  -2.27 & -1.18 & -1.9 & -2.05 & -2.27 & -0.62 & -1.805  & -2.30 \\
$\Sigma^{+}$ & 2.79 & 2.458 & 2.61 & 2.64 & & & & & 2.46 & 2.87 & & \\
$\Sigma^{0}$ & 0.839 & & & & &  0.11 & & & 0.65 &  0.76 & & \\
$\Sigma^{-}$ & -1.113 & -1.16 & -1.01 & -1.28 & & -1.13 & & & -1.16 & -1.48 & & \\
$\Sigma^{*+}$ & 2.877 & & 3.02 & 3.07 & 2.85 & & 2.56 & 2.54 & 2.63 &  1.27 & 3.028 & \\
$\Sigma^{*0}$ & 0.353 & & 0.30 & 0.08 & 0.09 & & 0.23 & 0.14 & 0.08 &  0.33 & 0.188 & \\
$\Sigma^{*-}$ & -2.171 & & -2.41 & -2.92 & -2.66 & & -2.10 & -2.19 & -2.43 & -1.88 & -2.015 & \\
$\Lambda^{0}$ & -0.606 & -0.613 & -0.59 & -0.60 & & -0.11 & & & -0.51 & -0.70 & & \\
\hline
\end{tabular}
\end{table*}

Table [\ref{tab:mmcompare}] gives comparison of present magnetic moment with those of different approaches. H. Dahiya et al. have presented octet and decuplet baryon magnetic moments in chiral quark model with configuration mixing and generalizing the Cheng-Li mechanism. Effective mass and screened charge scheme has been employed in ref. \cite{dhir} and both the results appear in the table. Light-cone sum rules \cite{hong} and lattice QCD \cite{flee} have also been employed for octet and decuplet magnetic moments. Recent study has focused on hyperonic medium at a finite temperature using chiral mean field approach \cite{harpreet}. Our results are under-predicted compared to experimental values of PDG by 0.5 and 0.4 respectively for $\Xi$ octet baryons, however, there are large variations in case of decuplet considering all the approaches. It is expected that experimental values for decuplet shall be the deciding factor. {\bf In case of $\Lambda^{0}$, the magnetic moment is nearly similar to those obatined by \cite{pdg}, \cite{harleen3} and \cite{flee}. For $\Sigma^{+}$ and $\Sigma^{-}$, results are within 0.5$\mu_{N}$ variation for all the approaches.} \\\\
Here, an effort has been made to determine the magnetic moment of low-lying negative parity state of $\Xi$ which has been inspired by studies based on N(1535). The hyperfine interactions between the constituent quarks induce linear combinations of two states with the same angular momentum value. The magnetic moment will now have contribution from spin as well as orbital angular momentum as \cite{liu}
\begin{equation}
\mu = \mu^{S} + \mu^{L} = \sum \frac{Q_{q}}{m_{q}}S_{q} + \sum \frac{Q_{q}}{2m_{q}}l_{q}
\end{equation}
The $J^{P}=\frac{1}{2}^{-}$, L=1 for $S=\frac{1}{2}$ and $S=\frac{3}{2}$, calculated resonance masses are 1886 MeV and 1894 MeV. These states are not experimentally established, however 1690 MeV is tentatively assigned by many predictions to this state. The physical eigenvalues will be 
\begin{equation}
|\Xi(1886)\rangle = cos\theta |^{2}P_{1/2}\rangle - sin\theta |^{4}P_{1/2}\rangle
\end{equation}
\begin{equation}
|\Xi(1894)\rangle = sin\theta |^{2}P_{1/2}\rangle + cos\theta |^{4}P_{1/2}\rangle
\end{equation}
From the approach of NCQM described in Ref. \cite{neetika}, we directly write the final wave-function with mixing for $\Xi^{0}(uss)$ as
\begin{equation}
(\frac{2}{9}\mu_{s} + \frac{1}{9}\mu_{u})cos^{2}\theta + (\mu_{s} + \frac{1}{3}\mu_{u})sin^{2}\theta - (\frac{8}{9}\mu_{s} - \frac{8}{9}\mu_{u})cos\theta sin\theta
\end{equation}
The mixing angle is taken as $\theta=-31.7$. The magnetic moment for $\Xi^{0}(1886)$ is obtained to be -0.695$\mu_{N}$ which is -0.99$\mu_{N}$ by \cite{neetika}. Similarly, for $\Xi^{-}(1886)$, magnetic moment is obtained as -0.193$\mu_{N}$ as compared to -0.315$\mu_{N}$ \cite{neetika}. The variation relies on the fact that resonance mass is a model dependent and the effective mass used in the calculated ultimately depends on resonance mass. 

\section{Transition Magnetic Moment and Radiative Decay Width}
Transition magnetic moment as well as radiative decay width are also important in the understanding of internal structure of baryon as well as magnetic and electric transitions. Many approach have been utilized for the study of radiative decay over years including some recent ones \cite{fayyazuddin}.{\bf  Here, $\Xi^{*0}  \rightarrow \Xi^{0}\gamma$ as well as $\Sigma^{*}  \rightarrow \Sigma\gamma$ have been studied using the effective mass obtained using hCQM approach. The generalized form for transition magnetic moment is \cite{harleen18},}
\begin{equation}
\mu(B_{\frac{3}{2}^{+}} \rightarrow B_{\frac{1}{2}^{+}}) = \langle B_{\frac{1}{2}^{+}}, S_{z}=\frac{1}{2}|\mu_{z}| B_{\frac{3}{2}^{+}}, S_{z}=\frac{1}{2} \rangle
\end{equation}
The spin-flavour wave function obtained in a similar way as above mentioned for  decuplet and octet $\Xi$ is
\begin{equation}
\frac{2\sqrt{2}}{3}(\mu_{u}^{eff}-\mu_{s}^{eff})
\end{equation}
The effective mass here is a geometric mean of those for spin $\frac{1}{2}$ and $\frac{3}{2}$. Our result comes out to be 2.378$\mu_{N}$ which is in good agreement with various models implemented in \cite{dhir,harleen18}.
The radiative decay width is obtained as \cite{kaushal},
\begin{equation}
\Gamma_{R}= \frac{q^{3}}{m_{p}^{2}}\frac{2}{2J+1}\frac{e^{2}}{4\pi}|\mu_{\frac{3}{2}^{+} \rightarrow \frac{1}{2}^{+}}|^{2}
\end{equation} 
where q is the photon energy, $m_{p}$ is the proton mass and J is the initial angular momentum giving $\Gamma_{R}$ = 0.214 MeV. The total decay width available from experiment is 9.1 MeV. Thus, the branching ratio $\frac{\Gamma_{R}}{\Gamma_{total}}$ is 2.35\% where the PDG data suggests $<3.7\%$. Thus, it is in accordance with other results as well as from experimental data. 
{\bf The radiative decay of $\Sigma$ baryons with transition magnetic moments are summarized in the following table [\ref{tab:radiative}]. The obtained results are consistent with experimental data from PDG as well as few other theoretical approaches. }

\begin{table*}
\centering
\caption{$\Sigma^{*}$ Radiative Decays}
\label{tab:radiative}
\begin{tabular}{cccc}
\hline
Decay & Wave-function & Transition moment(in $\mu_{N}$) & $\Gamma_{R}$(in MeV)\\
\hline
$\Sigma^{*0} \rightarrow \Lambda^{0}\gamma$ & $\frac{\sqrt{2}}{\sqrt{3}}(\mu_{u}-\mu_{d})$ & 2.296 & 0.4256 \\
$\Sigma^{*0} \rightarrow \Sigma^{0}\gamma$ & $\frac{\sqrt{2}}{3}(\mu_{u}+\mu_{d}-2\mu_{s})$ & 0.923 & 0.0246 \\
$\Sigma^{*+} \rightarrow \Sigma^{+}\gamma$ & $\frac{2\sqrt{2}}{3}(\mu_{u}-\mu_{s})$ & 2.204 & 0.1404 \\
$\Sigma^{*-} \rightarrow \Sigma^{-}\gamma$ & $\frac{2\sqrt{2}}{3}(\mu_{d}-\mu_{s})$ & -0.359 & 0.0037 \\
\hline
\end{tabular}
\end{table*}

\section{Conclusion}
{\bf The present work has been aimed at studying the strangeness -1 $\Lambda$, $\Sigma$ and -2 $\Xi$ light baryon owing to its limited data. The non-relativistic hypercentral Constituent Quark Model (hCQM) has been a tool for obtaining large number of resonance masses with a linear term. The results have been compared for with and without first order correction terms as well wherein a few MeV difference for low-lying states and upto the order of 30 MeV for higher excited states for $\Xi$. The octet and decuplet states have not been exclusively distinguished due to lack of required data. } \\\\
The mass-range have been compared to various theoretical approaches listed in section 3. The state-wise comparison is not possible because no approach has established spin-parity assignments. The overview of Tables \ref{tab:positive} and \ref{tab:negative} shows that the low-lying resonance masses are in good agreement among each other especially the four star states of PDG. For higher excited states, present work over-estimates the results compared to other models. The $\Xi$(1820) differs by 48 MeV from PDG mass. The $\Xi(2030)$ state with $J=\frac{5}{2}$ could find a place with either positive or negative parity a both have masses in that range. $\Xi$(1620) and $\Xi$(1690) are not obtained in present results. However, $\Xi$(1690) which is likely $\frac{1}{2}^{-}$ by BABAR Collaboration, this study calculates it as 1886 MeV. Few differences in results owes to the model dependent factors. {\bf  As for $\Lambda$(1405), hCQM could not establish the mass but predicts the other negative parity state with good agreement. The four star states for $\Sigma$ as well as $\Lambda$ agree to a good extent with many approaches too. } \\\\
The Regge trajectories based on some resonance mass are plotted. The principle quantum number n against the square of mass $M^{2}$ shows linear nature but fitted lines are not exactly parallel. The angular momentum J versus the square of mass $M^{2}$ plots also depicts the linearly of data points which validates our spin-parity assignments to a particular state as well as might be helpful in new experimental states. \\\\
The magnetic moment for spin $\frac{1}{2}$, $\Xi^{0}$ and $\Xi^{-}$ vary by 0.25$\mu_{N}$ and 0.19$\mu_{N}$ from PDG as well as other results. For spin $\frac{3}{2}$, $\Xi^{*0}$ and $\Xi^{*-}$, our result vary from nearly 0.2$\mu_{N}$ to 0.4$\mu_{N}$ compared to all the approaches.  The transition magnetic moment is obtained for $\Xi^{0}$ 1886 MeV of our spectra which differs by 0.30$\mu_{N}$ and $\Xi^{-}$ differs by 0.12$\mu_{N}$ from Ref \cite{neetika}. {\bf However, difference of magnetic moment follows due to difference of resonance masses used for the calculation. Similarly for $\Sigma^{+}$, $\Sigma^{-}$ and $\Lambda^{0}$ results are very well in accordance with other approaches and PDG. } \\\\
{\bf The transition magnetic moment has been obtained as 2.378$\mu_{N}$ which agrees with other models. Also, the radiative decay width branching fraction comes out to be 2.35\% in our case which is well within the PDG range of $<3.7\%$. The transition magnetic moments and radiative decay widths for $\Sigma$ falls similar to other results.  \\\\
Thus, with some agreements and some discrepancies, this study with large number of predicted resonances along with important properties, might be find helpful for upcoming experimental facilities like PANDA which is expected to intensively study the light strange baryons \cite{panda,barruca}. }

\section*{Acknowledgement}
Ms. Chandni Menapara would like to acknowledge the support from the Department of Science and Technology (DST) under INSPIRE-FELLOWSHIP scheme for pursuing this work.\\
  
%\section*{References}

\end{multicols}

\clearpage
\end{CJK*}
\end{document}